\documentclass[aps,pra,twocolumn,superscriptaddress,longbibliography]{revtex4-1}

\newcommand{\Er}{\ensuremath{{\cal E}_{\rm R}}}
\newcommand{\Eo}{\ensuremath{{\cal E}_{1,2,3}}}

\usepackage{amsmath}
\usepackage{amssymb}
\usepackage{amsfonts}
\usepackage{graphicx}
\usepackage[T1]{fontenc}
\usepackage{color}
\usepackage{textcomp}
\usepackage{lipsum}
\usepackage{braket}

\begin{document}

    \title{Four-wave mixing dynamics of a strongly coupled quantum-dot--microcavity system driven by up to 20 photons}

    \author{Daniel~Groll}
    \email{daniel.groll@uni-muenster.de}
    \affiliation{Institute of Solid State Theory, University of M\"{u}nster, 48149 M\"{u}nster, Germany}

    \author{Daniel~Wigger}
    \affiliation{Institute of Solid State Theory, University of M\"{u}nster, 48149 M\"{u}nster, Germany}
    \affiliation{Department of Theoretical Physics, Wroc\l{}aw University of Science and Technology, 50-370 Wroc\l{}aw, Poland}

    \author{Kevin~J\"urgens}
    \affiliation{Institute of Solid State Theory, University of M\"{u}nster, 48149 M\"{u}nster, Germany}

    \author{Thilo~Hahn}
    \affiliation{Institute of Solid State Theory, University of M\"{u}nster, 48149 M\"{u}nster, Germany}

    \author{Christian~Schneider}
    \affiliation{Technische Physik, University of W\"{u}rzburg, 97074 W\"{u}rzburg, Germany}

    \author{Martin~Kamp}
    \affiliation{Technische Physik, University of W\"{u}rzburg, 97074 W\"{u}rzburg, Germany}

    \author{Sven~H\"ofling}
    \affiliation{Technische Physik, University of W\"{u}rzburg, 97074 W\"{u}rzburg, Germany}
    \affiliation{SUPA, School of Physics and Astronomy, University of St. Andrews, St. Andrews KY16 9SS, UK}

    \author{Jacek~Kasprzak}
    \affiliation{Universit\'{e} Grenoble Alpes, CNRS, Grenoble INP, Institut N\'{e}el, 38000 Grenoble, France}

    \author{Tilmann~Kuhn}
    \affiliation{Institute of Solid State Theory, University of M\"{u}nster, 48149 M\"{u}nster, Germany}

    \begin{abstract}       
       The Jaynes-Cummings (JC) model represents one of the simplest ways in which single qubits can interact with single photon modes, leading to profound quantum phenomena like superpositions of light and matter states. One system, that can be described with the JC model, is a single quantum dot embedded in a micropillar cavity. In this joint experimental and theoretical study we investigate such a system using four-wave mixing (FWM) micro-spectroscopy. Special emphasis is laid on the dependence of the FWM signals on the number of photons injected into the microcavity. By comparing simulation and experiment, which are in excellent agreement with each other, we infer that up to $\sim20$ photons take part in the observed FWM dynamics. Thus we verify the validity of the JC model for the system under consideration in this non-trivial regime. We find that the inevitable coupling between the quantum dot exciton and longitudinal acoustic phonons of the host lattice influences the real time FWM dynamics and has to be taken into account for a sufficient description of the quantum dot-microcavity system. Performing additional simulations in an idealized dissipation-less regime, we observe that the FWM signal exhibits quasi-periodic dynamics, analog to the collapse and revival phenomenon of the JC model. In these simulations we also see that the FWM spectrum has a triplet structure, if a large number of photons is injected into the cavity.
    \end{abstract}


    \maketitle

\section{Introduction}
Recently, the strong coupling regime of cavity quantum
electrodynamics received renewed attention in the context of a
coupled quantum dot (QD) microcavity system, clearly demonstrating
vacuum Rabi oscillations, anti-crossing of the polariton branches
and spectral signatures of multi-photon transitions in an ultra
low-loss open-cavity setup~\cite{najer2019gated}. The theoretical
description of such systems has a long lasting history, starting off
with the formulation of the Jaynes-Cummings (JC)
model~\cite{jaynes1963comparison}, representing one of the simplest
ways to couple a two-level system (TLS) and a single photon mode.
The significance of this model lies in its universality, meaning
that it can be used to accurately describe the dynamics of seemingly
different physical systems. Examples range from Rydberg atoms in
microwave
cavities~\cite{brune1996quantum,raimond2001manipulating,birnbaum2005photon},
via superconducting qubits coupled to on-chip microwave
cavities~\cite{fink2008climbing} to QDs embedded in optical
resonators~\cite{faraon2008coherent,kasprzak2010up,volz2012ultrafast,kasprzak2013coherence,cygorek2017nonlinear,najer2019gated}.
The JC system assembles many features that are inherently quantum in
nature. For example, if the TLS is in the excited state and the
photon mode in its vacuum state, one can observe the coherent exchange of
excitation between the TLS and a single photon, known as vacuum Rabi
oscillations~\cite{brune1996quantum,najer2019gated}. Furthermore the
effective TLS-photon coupling strength scales with the number of
photons in a nonlinear fashion, giving rise to a characteristic
spectrum, known as the JC ladder. It consists of doublets of energy
levels, called the rungs of the ladder. Its specific spectral
properties lead to important features like photon
blockade~\cite{birnbaum2005photon,faraon2008coherent}, where an
excitation of the JC system with a certain resonance frequency
blocks subsequent excitations with the same frequency. It also
permits to build all-optical photon switches, that could be used for
long-distance optical communication and quantum
computation~\cite{volz2012ultrafast}.

Verifying the JC model can be achieved by observing its spectrum
directly or indirectly. Historically, this has first been achieved
in atomic physics~\cite{brune1996quantum}. Using a spectroscopic
pump-probe technique the level structure of the JC ladder could also
be resolved in superconducting qubit systems up to the second rung
of the ladder~\cite{fink2008climbing}. In the context of QDs coupled
to optical cavities the four-wave mixing (FWM)
microscopy~\cite{langbein2010coherent} was successfully applied to
observe the influence of the first two rungs of the JC ladder on the
dynamics of the measured FWM
signal~\cite{kasprzak2010up,kasprzak2013coherence}.

In this study, we investigate a strongly coupled QD-microcavity
system using the coherent FWM micro-spectroscopy technique.
Specifically, we concentrate on the influence of the number of
photons present in the cavity on the observed FWM dynamics. By
comparison with predictions for the FWM signal of the QD-microcavity
system, we verify the validity of the JC model. From the simulation
we can infer, that after a pulsed optical excitation up to 20
photons interact with the QD exciton. For an accurate description of
the entire system, one has to take into account the inevitable
coupling of the exciton to the phonons of the QD's host lattice.
There are several approaches in the literature to study the dynamics
of this particular
system~\cite{roy2011influence,nazir2016modelling,cygorek2017nonlinear}.
Here, we use a Markovian Lindblad approach in the polaron frame~\cite{breuer2002theory, nazir2016modelling}.
\section{Experiment}
\label{sec:experiment}
In the experiments, we employed a sample containing arrays of
micropillar cavities~\cite{wigger2018rabi}. The pillars were
etched down from a planar structure, made of two distributed Bragg
reflectors enclosing a spacer which contains a layer of InAs QDs.
The dots formed at the antinode of the photon field and show large
oscillator strengths~\cite{reithmaier2004strong}, owing to the
increased indium content. The resonator studied here, displaying a
quality factor of around 22000 and diameter of 1.8\,\textmu m, operates in the strong coupling regime. This was verified by observing a
polaritonic anti-crossing of the non-resonantly excited
photoluminescence, when temperature-tuning the QD exciton (X)
transition across the cavity (C). The sample is kept in a He-flow
optical cryostat and the X-C resonance was found at $T=23\,$K.

\begin{figure}
    \centering
    \includegraphics[width=\linewidth]{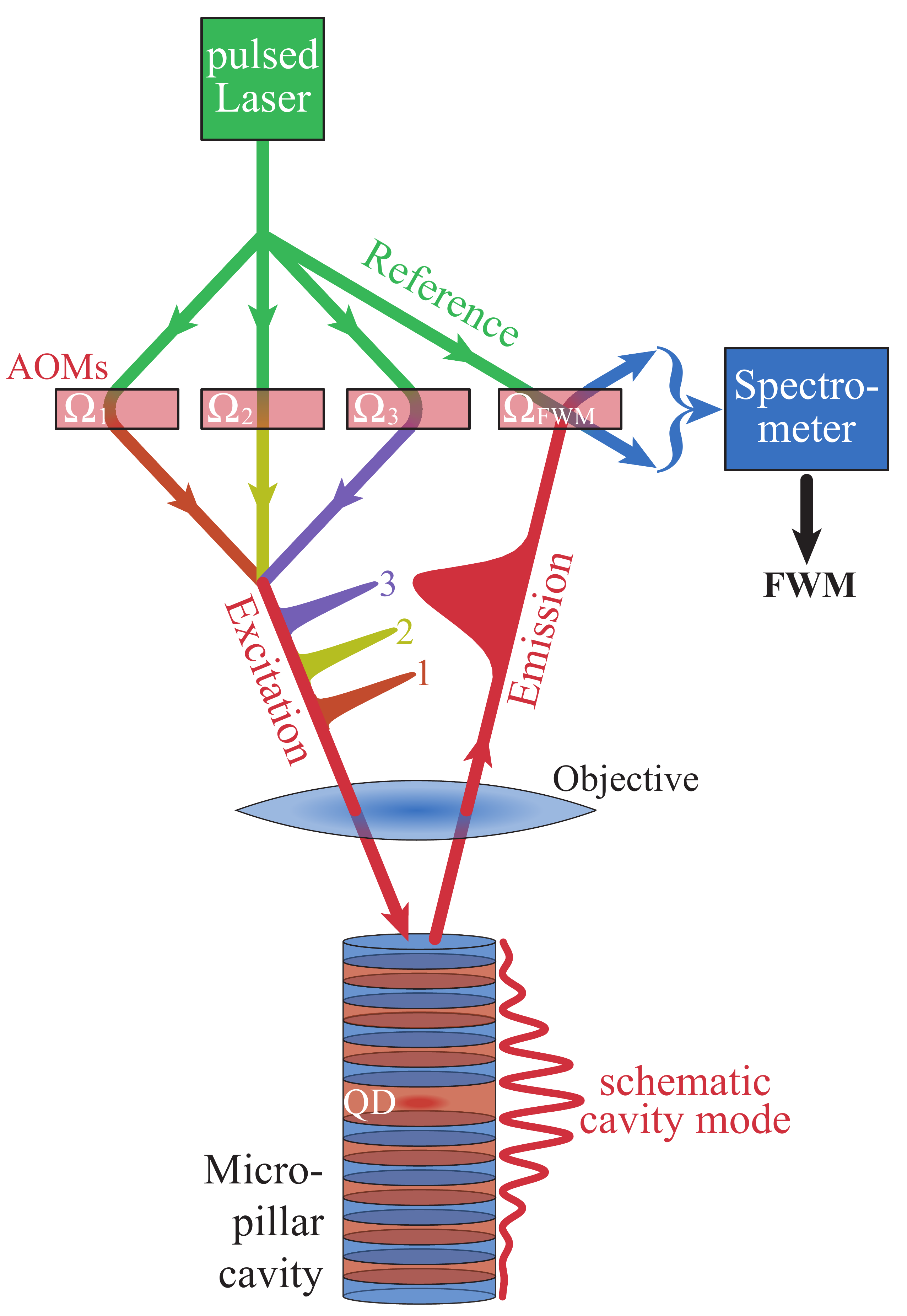}
    \caption{Schematic view of the experimental methodology.
    A triplet of frequency-shifted, short laser pulses is focussed with a microscope objective onto the top-facet of a micropillar cavity.
    The reflected light is collected with the same objective and interfered with the reference beam shifted to the FWM frequency.
    The FWM spectral interference is then measured on a CCD camera, installed at the output of the imaging spectrometer.}
    \label{fig:exp}
\end{figure}

Using an external microscope objective, we tightly focus three laser
pulses $\Eo$ onto a top facet of a pillar, as schematically shown in Fig.~\ref{fig:exp}, to mode-match the excitation with the fundamental
(two-dimensional Gaussian) transverse mode. Pulses of an initial
duration of around 500\,fs and a central wavelength around 946\,nm, are stretched to the lifetime of the cavity mode inside the resonator, i.e. approx 13\,ps. The
polaritonic FWM signal is therefore resonantly driven through the
photonic component. It is then emitted principally in upward
direction, collected via the same objective and directed towards the
spectrally-resolved detection with an imaging spectrometer and a CCD
camera.

Owing to the microscopy configuration of the experiment, requiring
co-linear propagation of $\Eo$, FWM is measured in a heterodyne
detection scheme. As depicted in Fig.\,\ref{fig:exp}, the primary
Ti:Sapphire laser beam is split and each of the three excitation components is
phase-shifted using acousto-optic modulators (AOMs) operating at
distinct radio-frequencies $\Omega_{1,2,3}$. While separated, the
excitation beams can acquire time-delays, $\tau_{12}$ and
$\tau_{23}$, adjustable with mechanical delay stages. FWM is
detected by interfering the emission with the reference field
$\Er$ in a spectrometer. Both fields are modulated with another AOM operating at the frequency carrying
the FWM response $\Omega_{\rm FWM}=\Omega_3+\Omega_2-\Omega_1$. Note, that the reference pulse is stretched in the spectrometer by the response function, discussed later.
From the stationary spectral-interference of the pulse trains we obtain the amplitude
and phase of the signal by applying spectral interferometry. To
measure the coherence dynamics of the JC polariton system, we vary
$\tau_{12}$, while keeping $\tau_{23}=0$ fixed.

\section{Theory}
\subsection{Modeling the QD-microcavity system}
\label{sec:model}
The system we want to describe consists of three parts, the QD exciton, the microcavity and the longitudinal acoustic (LA) phonons. An external laser field in the form of short laser pulses injects photons into the cavity through the top mirror. The frequency of the lowest lying cavity mode is taken to be in close resonance with the lowest lying bright exciton energy. By choosing an accurate polarization for the laser field, which determines the polarization of the cavity mode, we can make the usual assumption that the QD exciton can be described as a two-level system~\cite{hohenester2007quantum}. The cavity mode is modeled by a single harmonic oscillator. The assumption here is that the external laser pulse spectrum is centered around the frequency of the lowest lying confined mode. The coupling between the cavity mode and the QD exciton is treated within the JC model~\cite{jaynes1963comparison}. In doing so we assume that the exciton states have a well defined parity, that the dipole approximation holds and that we can neglect counterrotating terms in the Hamiltonian, i.e. we perform a rotating wave approximation in the exciton photon coupling. The latter is justified since we assume that cavity and exciton frequencies are in or close to resonance with each other. It is known that in InAs/GaAs QDs, if electron and hole are not spatially separated, e.g. by static electric fields, the dominant coupling mechanism of phonons to the exciton leading to dephasing is the deformation potential coupling to LA phonons~\cite{krummheuer2002the}. The total Hamiltonian of the system is given by
    \begin{align}\label{eq:H_tot}
    H=&\hbar\omega_X X^{\dagger}X+\hbar\omega_c a^{\dagger}a+\hbar g_{\rm JC}\left(aX^{\dagger}+a^{\dagger}X\right)\notag\\
    &+\sum_{\bf q}\hbar\omega_qb_{\bf q}^{\dagger}b_{\bf q}^{}+X^{\dagger}X\sum_{\bf q}\hbar g_q\left(b_{\bf q}^{\dagger}+b_{\bf q}^{}\right)\ ,
    \end{align}
where $X=\ket{G}\bra{X}$ is the exciton annihilation operator, with $\ket{G}$ and $\ket{X}$ being the excitonic ground and excited state, respectively. $a$ and $a^\dagger$ are the boson annihilation and creation operators of the cavity photons and $b_{\bf q}^{}$ and $b_{\bf q}^\dagger$ are the boson annihilation and creation operators for the phonons with wave vector ${\bf q}$. The exciton and cavity frequency are $\omega_X$ and $\omega_c$, respectively. $\omega_q=c_s|{\bf q}|$ are the LA phonon frequencies in the Debye model, where $c_s$ is the speed of sound. The coupling constant for the exciton-photon coupling is denoted by $g_{\rm JC}$ and the deformation potential coupling constants are given by~\cite{lueker2017phonon}
    \begin{equation}\label{eq:g_q}
    g_q=\sqrt{\frac{|{\bf q}|^2}{2\rho_m\hbar V\omega_q}}\left(D^ee^{-\frac{1}{4}|{\bf q}|^2a_e^2}-D^he^{-\frac{1}{4}|{\bf q}|^2a_h^2}\right)\ .
    \end{equation}
Here, $\rho_m$ is the mass density of the host material of the QD. $D^{e/h}$ are the deformation potential constants and $a_{e/h}$ the extensions of the wave function of electron and hole, respectively. $V$ is the normalization volume of the phonons. This leads to the corresponding spectral density
    \begin{align}
    J(\omega)&=\sum_{\bf q}g_q^2\delta(\omega-\omega_{|\bf q|})=\frac{V}{2\pi^2 c_s}\left(\frac{\omega}{c_s}\right)^2g_{|\omega/c_s|}^2 \label{eq:specdens}\\
    &=\frac{1}{4\pi^2\hbar\rho_m}\frac{\omega^3}{c_s^5}\left[D_e e^{-\frac14\left(\omega a_e/c_s\right)^2}-D_h e^{-\frac{1}{4}\left(\omega a_h/c_s\right)^2}\right]^2\ .\notag
    \end{align}
    For simplicity we assume a spherical geometry of the QD and model a realistic spectral density by choosing the parameters $a_e$ and $a_h$ independently from each other~\cite{lueker2017phonon}. The QD and phonon parameters chosen in this work are $c_s=5122$\,m\,s$^{-1}$, $\rho_m=5.317$\,g\,cm$^{-3}$~\cite{blakemore1982semiconducting}, $D_e=7\,$eV, $D_h=-3.5\,$eV~\cite{selbmann1996coupled}, $a_e=7\,$nm, $a_h=1.5\,$nm~\cite{wigger2018rabi}.

    The master equation approach, that will be used to treat the interaction between the coupled exciton-photon system and the phonons, relies on a perturbation expansion in terms of the respective interaction Hamiltonian.  Instead of directly treating the exciton-phonon interaction in Eq.~\eqref{eq:H_tot} as the perturbation, the unitary polaron transformation can be performed, which diagonalizes the Hamiltonian in the absence of the exciton-photon coupling ($g_{\rm JC}=0$)~\cite{nazir2016modelling}. This corresponds to a transformation from the exciton frame to the polaron frame. Then a residual polaron-phonon-cavity interaction is treated as the perturbation, see Eq.~\eqref{eq:H_I_P}. In the appendix in Fig.~\ref{fig:11} (a) and (b), simulations of vacuum Rabi oscillations, obtained within the master equation approach in the polaron frame, are compared to correlation expansion calculations, showing good agreement for the parameters relevant in this work. Further details on the polaron transformation, the derivation of the Lindblad master equation and the influence of the approximations used in the Lindblad approach are discussed in the appendix.

    In the polaron frame we still have a JC Hamiltonian governing the dynamics of the polaron, coupled to the cavity mode. This Hamiltonian is given by (see also Eq.~\eqref{eq:H_S_P})
    \begin{equation}\label{eq:H_S_P_main}
    H_S^P=\hbar\omega_c a^{\dagger}a+\hbar\tilde{\omega}_X X^{\dagger}X+\hbar\tilde{g}\left(aX^{\dagger}+a^{\dagger}X\right)\,.
    \end{equation}
    $\tilde{g}$ is the polaron photon coupling and $\widetilde{\omega}_X$ the polaron shifted exciton frequency. The spectrum of $H_S^P$ is called the JC ladder. It contains the ground state $\ket{G,0}$ with zero photons and no polaron present. The ground state fulfills $H_S^P\ket{G,0}=0$. The rest of the JC ladder consists of doublets of eigenstates $\ket{n,\pm}$ for every $n>0$. $n$ is called the rung number of the JC ladder. The eigenstates $\ket{n,\pm}$ in the $n$-th rung are superpositions of the states $\ket{G,n}$ and $\ket{X,n-1}$. The second entry in the ket vectors denotes the number of photons present in the cavity. The energies of the eigenstates of the $n$-th rung are given by $E_n^{\pm}=E_n^0\pm \frac{1}{2}\hbar\widetilde{\Omega}_n$ with the Rabi splitting of the n-th rung $\hbar \widetilde{\Omega}_n$. In the case of vanishing detuning $\delta=\omega_c-\tilde{\omega}_X=0$, they are given by
    \begin{equation}
    E_n^{\pm}|_{\delta=0}=\hbar n\omega_c\pm\hbar\tilde{g}\sqrt{n}\,.
    \end{equation}
    The JC ladder is depicted in Fig.~\ref{fig:2} for the case $\delta=0$. It is shown there how the energy splitting in each rung grows with $\sqrt{n}$.

    The Lindblad master equation, describing transitions in the JC ladder due to the presence of phonons and spontaneous emission of photons is given by (see also Eq.~\eqref{eq:Lindblad_full_appendix})
    \begin{align}\label{eq:Lindblad_full}
    \hbar\frac{\text{d}}{\text{d}t}\rho_S^P(t)=&-i\left[H_S^P+H_{LS},\rho_S^P(t)\right]\\
    &+\mathcal{D}\left[\rho_S^P(t)\right]+\mathcal{D}_a\left[\rho_S^P(t)\right]+\mathcal{D}_X\left[\rho_S^P(t)\right]\notag\ .
    \end{align}
    It describes the time evolution of the reduced density matrix of the JC system in the polaron frame. The commutator term on the right hand side describes the unitary time evolution. This time evolution is governed by the Hamiltonian of the JC system itself, $H_S^P$ as given in Eq.~\eqref{eq:H_S_P_main}. Additionally, the Lamb shift Hamiltonian $H_{LS}$ appears; for further details see Eq.~\eqref{eq:H_LS}. This describes residual energy renormalizations in the JC system, induced by the phonon bath at temperature $T$, which are not already captured by the polaron transform. The phonon dissipator $\mathcal{D}$, as given by Eq.~\eqref{eq:lindblad_D} with Eqs.~\eqref{eq:A_P}, describes transitions in the JC ladder, induced by the phonons. It contains terms describing spontaneous and induced emission, as well as absorption processes. The interaction between exciton and phonons, as given in Eq.~\eqref{eq:H_tot}, preserves the number of photons present in the cavity and the occupation of the exciton and thus also of the polaron. Therefore the phonons induce only transitions within a rung of the JC ladder and no inter-rung transitions. This is depicted in Fig.~\ref{fig:2} by the red arrow. $\mathcal{D}_a$ and $\mathcal{D}_X$, as given in Eq.~\eqref{eq:dissipator}, are called cavity and polaron dissipators and describe photon losses through the mirrors and spontaneous emission into unconfined modes with rates $\gamma_a$ and $\gamma_X$, respectively. They induce transitions from the $n$-th to the $(n-1)$-th rung. This is depicted in Fig.~\ref{fig:2} by the green arrow.
    
\subsection{Modeling cavity feeding and FWM signal extraction}
\begin{figure}[t]
	\centering
	\includegraphics[width=0.6\linewidth]{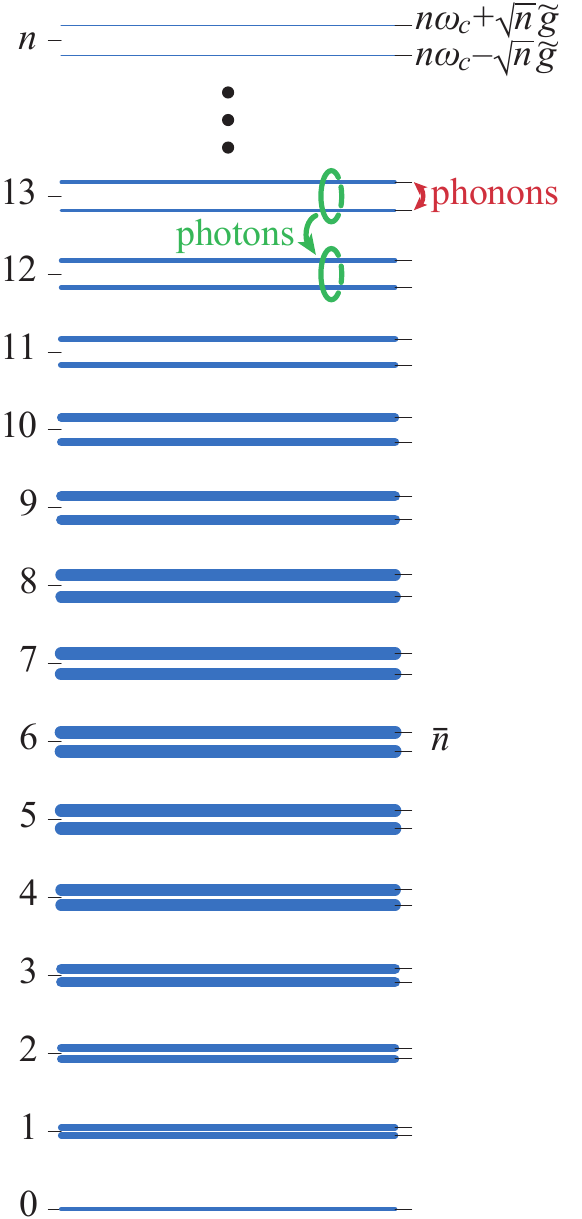}
	\caption{Schematic picture of the JC ladder for $\delta=\omega_c-\tilde{\omega}_X=0$. $n$ labels the rung of the doublets. The splitting within each doublet increases with $\sqrt{n}$. The thickness of the lines illustrates the occupation of the states after a pulsed excitation from the ground state $\ket{G,0}$, generating a coherent state. The red arrow indicates transitions inside a rung, induced by the LA phonons. The green arrow indicates transitions between adjacent rungs by spontaneous emission of photons from the cavity and the polaron.}
	\label{fig:2}
\end{figure}

The cavity mode is excited by short laser pulses with a duration in the sub-ps range. The interaction of an external classical laser field ${\bf{E}}_c$ with the cavity mode through the top mirror is modeled by a tunneling interaction of the form~\cite{kasprzak2010up}
    \begin{equation}
    V_c(t)={\boldsymbol \mu}_c\cdot{\bf E}_c(t)a^{\dagger}+h.c.\ ,
    \end{equation}
    where ${\boldsymbol \mu}_c$ is the effective dipole matrix element of the cavity mode. As explained in Sec.~\ref{sec:experiment}, the pulses from the laser are much faster than the typical timescale of the cavity, such that the laser field is taken to be a sequence of delta pulses
    \begin{equation}
    {\bf E}_c(t)=\sum_j {\bf E}_c^j \delta(t-t_j)\ .
    \end{equation}
We define the pulse areas of the pulses as
    \begin{equation}\label{eq:def_pulse_area}
    \Theta_j=\frac{2{\boldsymbol \mu}_c\cdot {\bf E}_c^j}{\hbar}\ ,
    \end{equation}
such that the interaction takes the form
    \begin{equation}\label{eq:pulse_interaction}
    V_c(t)=\sum_j\hbar\frac{\Theta_j}{2}a^{\dagger}\delta(t-t_j)+h.c.\ .
    \end{equation}
Between the pulses, the system evolves freely according to the Hamiltonian in Eq.~\eqref{eq:H_tot}. A single pulse at time $t_j$ transforms the density matrix of the system $\rho$ unitarily according to
    \begin{equation}\label{eq:cavity_displacement}
    \rho\rightarrow D(\alpha_j)\rho D^{\dagger}(\alpha_j)
    \end{equation}
with $D(\alpha_j)=\exp(\alpha_ja^{\dagger}-\alpha_j^*a)$ being the displacement operator of the cavity mode with the coherent amplitude $\alpha_j=-i \Theta_j/2$~\cite{glauber1963coherent}. This can be derived by considering the full time evolution operator of the system Hamiltonian in Eq.~\eqref{eq:H_tot} plus the interaction with external laser pulses in Eq.~\eqref{eq:pulse_interaction} and taking the limit of an infinitesimal time evolution around the pulse excitation at time $t_j$. This is equivalent to neglecting the dynamics of the system and especially the exciton-photon interaction, as described by Eq.~\eqref{eq:H_tot}, during the infinitesimal pulse duration. Thus, if the cavity is in the vacuum state before the pulse, it is in a coherent state afterwards. For a coherent state the occupation of the photon states has a Poissonian distribution with an expectation value of~\cite{glauber1963coherent}
    \begin{equation}\label{eq:n}
    \overline{n} = |\alpha_j|^2=\Theta_j^2/4\ .
    \end{equation}
This situation is depicted in Fig.~\ref{fig:2} by the thickness of the lines marking the energy levels. If multiple pulses arrive at the same time, according to Eq.~\eqref{eq:pulse_interaction} their pulse areas and thus also their coherent amplitudes add up. With the product rule for displacement operators~\cite{glauber1963coherent}
    \begin{equation}\label{eq:multiple_pulses}
    D(\alpha+\beta)=D(\alpha)D(\beta)\exp[ i\text{Im}(\alpha\beta^\ast)]
    \end{equation}
the interaction with multiple pulses at the same time can be written as a sequence of single pulse interactions. The ordering of these single pulse interactions does not matter, since the phase factor in Eq.~\eqref{eq:multiple_pulses} cancels in the transformation of the density matrix in Eq.~\eqref{eq:cavity_displacement}.

In this study we are especially interested in FWM spectroscopy signals. On the theoretical side, this is modeled by assigning  a phase $\phi_j$ to the $j$-th pulse~\cite{wigger2017exploring}
    \begin{equation}
    {\bf E}_c^j\rightarrow {\bf E}_c^j\exp(-i\phi_j)\ ,\qquad \Theta_j\rightarrow\Theta_j\exp(-i\phi_j)\ .
    \end{equation}
In the experiment, a heterodyne detection scheme is employed, where the exciting pulses are frequency shifted by acousto-optical modulators~\cite{langbein2010coherent}. The phases $\phi_j$ take the role of the phase shifts that are induced by these radio-frequency shifts, as explained in Sec.~\ref{sec:experiment}. After the interaction with the pulses, the density matrix of the system contains contributions of all different phase combinations. Using this phase information one can select specific parts of the entire density matrix by integrating over the phases $\phi_j$. In the experiment this is achieved by using pulse trains and accumulating the single events. This phase selection process is defined by the transformation
    \begin{equation}\label{eq:phase_selection}
    \rho\rightarrow \frac{1}{(2\pi)^{N_p}}\int\limits_0^{2\pi}\rho\prod\limits_{j=1}^{N_p}\exp(if_j\phi_j)\text{d}\phi_j\ .
    \end{equation}
This selects the parts of the density matrix proportional to
    \begin{equation}
    \prod\limits_{j=1}^{N_p}\exp(-if_j\phi_j)
    \end{equation}
    setting all other terms to zero. $f_j$ are integers and $N_p$ is the number of pulses. In general the density matrix will no longer be hermitian and normalized after the phase selection, as some of its elements are set to zero. However, due to the linearity of the time evolution between the pulses and of the pulse interactions themselves, we can perform the phase integration directly after each pulse without spoiling the systems dynamics. Note that due to the phase selection process part of the information on the system is lost and strictly speaking the phase selected density matrix is no density matrix anymore. However, for simplicity we will continue calling it density matrix.

In the FWM experiment the system is excited by three laser pulses and the electric field, which is emitted by the cavity through the top mirror in axial direction is measured after the last pulse. The relevant observable is the electric field of the cavity and due to the phase selection this is proportional to the cavity field $\braket{a}$~\cite{kasprzak2010up,kasprzak2013coherence}. We will categorize the FWM signals according to the notation
    \begin{equation}
    \phi_{\rm FWM}=f_1\phi_1+f_2\phi_2+f_3\phi_3\,,
    \end{equation}
which means that we phase select the part of the cavity field proportional to
    \begin{equation}
        \left< a\right>_{\rm FWM}\sim e^{-i(f_1\phi_1+f_2\phi_2+f_3\phi_3)}\ .
    \end{equation}
To obtain a non-vanishing cavity field, we must satisfy $\sum_j f_j=1$. The method is called FWM only if the absolute values of the phase factors add up to $3$. In the case of weak laser pulses, we measure the third order nonlinear polarization~\cite{langbein2010coherent}. The concept can be generalized to $N$ wave mixing, if we choose $\sum_j |f_j|=N-1$.

\section{Results}
We consider the case of three pulse FWM, i.e., the system is excited by three short laser pulses with pulse areas $\Theta_j$ with $j=1,2,3$ and relative delays $\tau_{ij}$. A positive delay $\tau_{ij}$ means, that pulse $i$ excites the system before pulse $j$. Likewise, for a negative $\tau_{ij}$, pulse $j$ excites the system before pulse $i$. Initially, the system is taken to be in the ground state $\rho=\ket{G,0}\bra{G,0}$.

The investigated FWM signals carry the phase relation $\phi_{\rm FWM}=-\phi_1+\phi_2+\phi_3$. We consider zero delay $\tau_{23}=0$ between pulses 2 and 3 throughout the paper. It has to be emphasized that this is in general not the same as degenerate FWM with $\phi_{\rm FWM}=-\phi_1+2\phi_2$. The signals are only the same in the low pulse area regime of pulse 2 and 3, where one can restrict the signals to the lowest order contribution of the pulse transformation from Eq.~\eqref{eq:cavity_displacement}, that survives the phase selection process in Eq.~\eqref{eq:phase_selection}.

The timescale, that counts from the last pulse excitation is called real time $t$. This time evolution of the system can be observed by measuring the FWM signal with a spectrometer. The spectrometer response function is given in Ref.~\cite{jakubczyk2016impact} and is in the simulation multiplied with the real time FWM amplitude to account for the finite spectral resolution. The reference pulse of the FWM setup arrives 3~ps before the last pulse excitation.

The laser intensity of each pulse is denoted by $P_j$ and due to the linear relationship between pulse area and electric field from Eq.~\eqref{eq:def_pulse_area} we connect pulse area and laser intensity via
    \begin{equation}
    \Theta_j=c\sqrt{P_j}\ ,
    \end{equation}
where $c$ is a proportionality factor, which is determined by fitting the theoretical calculations to the experiment. The parameters, for which simulation and experiment agree well and which are used throughout the paper, unless stated otherwise, are $\hbar\tilde{g}=35$~\textmu eV, $T=23$~K, $\delta=0$, $\hbar\gamma_a=50$~\textmu eV, $\hbar\gamma_X=2$~\textmu eV, $c=2.5\pi\ \sqrt{\text{\textmu W}}^{-1}$. The phonon and QD parameters are given in section \ref{sec:model}. All calculations are performed in the frame rotating with the cavity frequency $\omega_c$.
\subsection{Pulse area dependent real time dynamics}
\subsubsection{Comparison to experiment}
\begin{figure}[t]
    \centering
    \includegraphics[width=\linewidth]{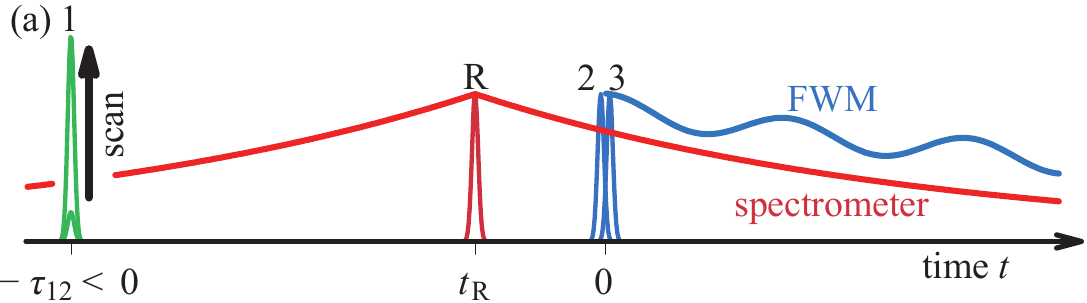}\\
    \includegraphics[width=\linewidth]{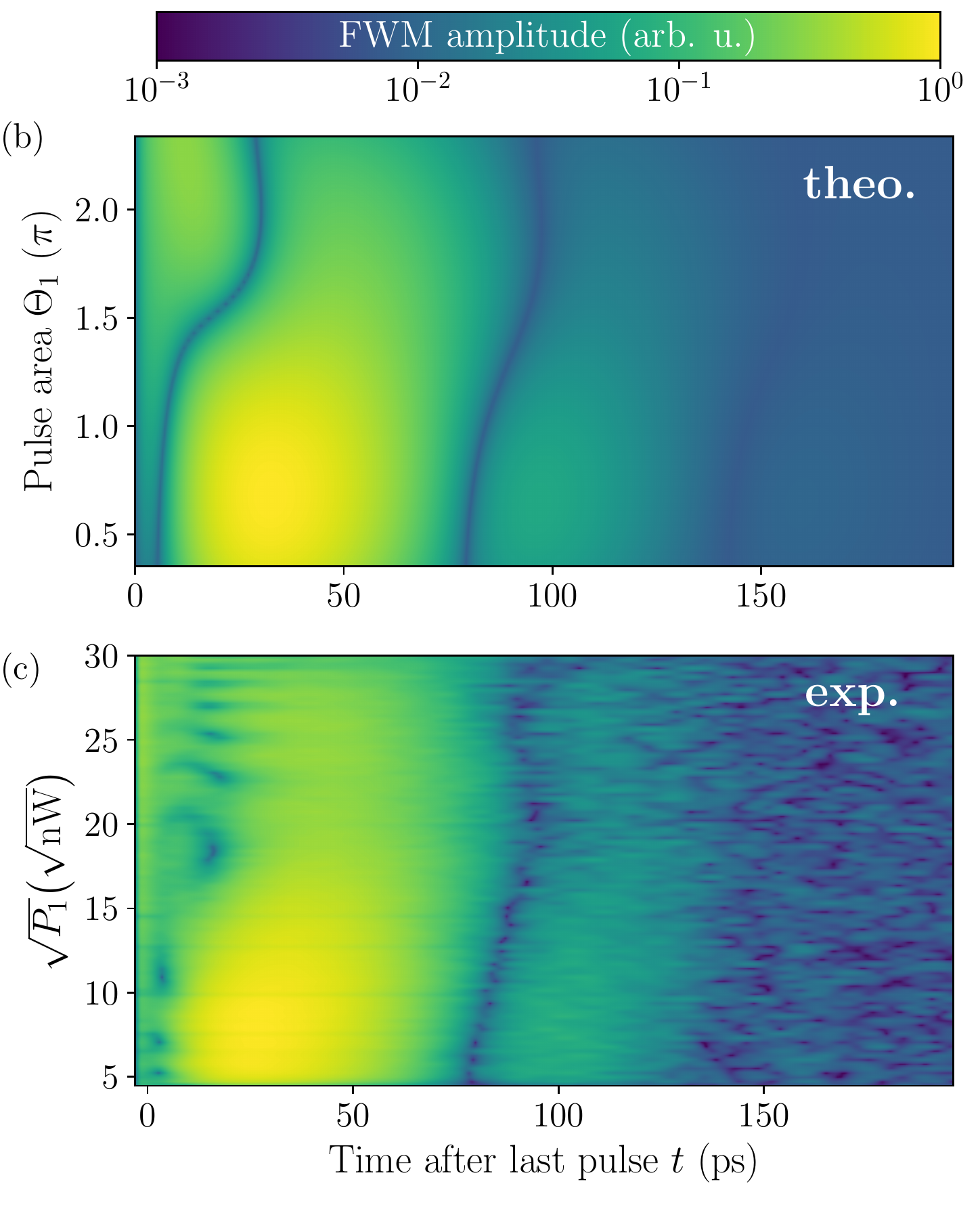}
    \caption{(a) Schematic picture of the experiment. Pulse 1 excites the system first, followed by pulse 2 and 3. The delay between pulse 1 and pulses 2 and 3 is $\tau_{12}=12$\,ps. Pulse 2 and 3 excite the system at the same time, i.e., $\tau_{23}=0$. The reference pulse R arrives 3\,ps before pulse 2 and 3 excite the system. (b) and (c) Simulated and measured real time dynamics of the FWM amplitude as a function of the pulse area $\Theta_1$ and laser field amplitude $\sqrt{P_1}$, respectively. The laser intensities of pulse 2 and 3 are $P_2=P_3=40\,$nW.}
    \label{fig:3}
\end{figure}

First we consider the real time dynamics of the FWM amplitude at different laser intensities in the experiment, respectively pulse areas in the simulation. The intensities of pulse 2 and 3 are set to $P_2=P_3=40$~nW, which corresponds to pulse areas $\Theta_2=\Theta_3=\pi/2$. In the theoretical calculations, the FWM amplitude corresponds to the phase filtered absolute value of the cavity field $|\braket{a}_{\rm FWM}|$. The ordering of the pulses for a delay of $\tau_{12}=12$~ps is schematically shown in Fig.~\ref{fig:3}(a). The reference pulse is denoted by R. The simulated and measured real time dynamics of the FWM amplitude are displayed in Fig.~\ref{fig:3}(b) and (c) in false colors as functions of the field amplitude $\sqrt{P_1}$ and pulse area $\Theta_1$ of pulse~1, respectively. A constant background was added in the simulation, that was retrieved by averaging the measured signal for $t>150$~ps.

\begin{figure}[t]
	\centering
	\includegraphics[width=\linewidth]{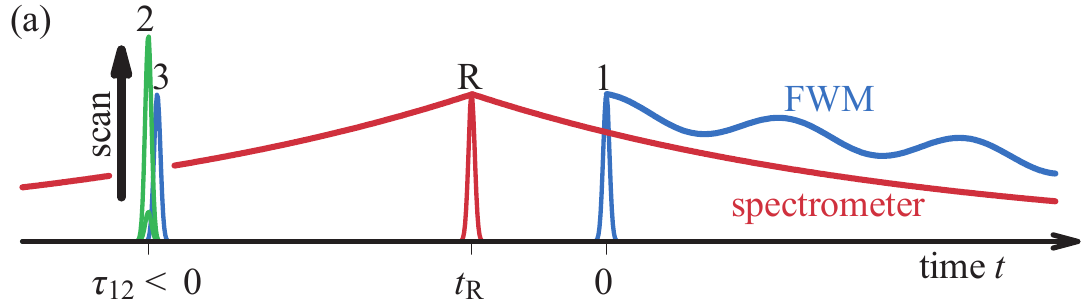}\\
	\includegraphics[width=\linewidth]{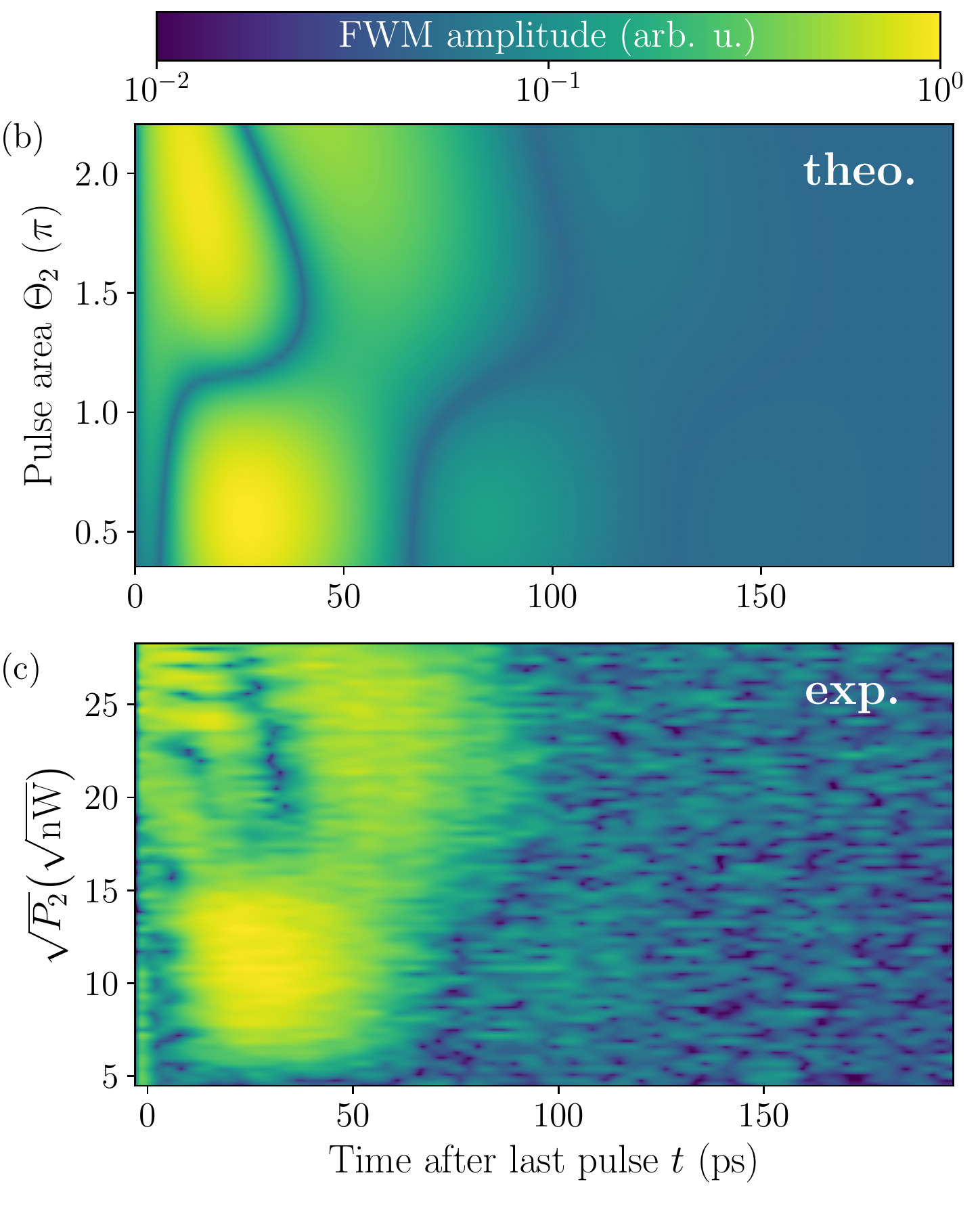}
	\caption{(a) Schematic picture of the experiment. Pulses 2 and 3 excite the system first, followed by pulse 1. The delay between pulse 1 and pulses 2 and 3 is $\tau_{12}=-10$\,ps. Pulse 2 and 3 excite the system at the same time, i.e., $\tau_{23}=0$. The reference pulse R arrives 3\,ps before the system is excited by pulse 1. (b) and (c) Simulated and measured real time dynamics of the FWM amplitude as a function of the pulse area $\Theta_2$ and laser field amplitude $\sqrt{P_2}$, respectively. The laser intensities of pulse 1 and 3 are $P_1=P_3=40\,$nW.}
	\label{fig:4}
\end{figure}

In the simulation, displayed in Fig.~\ref{fig:3}(b), we observe damped oscillations of the FWM amplitude for all pulse areas. We can use the minima of the signal, represented by blue colors, as reference points in the plot. We find that the positions of the minima shift when increasing the pulse area. For pulse areas up to $\Theta_1\approx 1.5\pi$, the first minimum appears roughly at the same position, yielding an almost vertical line for the first minimum. For larger pulse areas the position of this minimum shifts towards later times $t$. A similar minimum line appears at $t\approx 80$~ps with a slightly different bending depending on $\Theta_1$. Later in Fig.~\ref{fig:7} we will see that the shape of the minimum line is periodically repeated from the second minimum line on, which is not visible here due to the background.

In the experiment in Fig.~\ref{fig:3}(c) we essentially observe the same damped oscillations and shifts of the minima with increasing pulse intensity. The only clear difference appears at small times during the first minimum. What  was a continuous minimum line in the simulation is here interrupted by an artifact of the experiment, such that the intensity forms an almost periodic modulation in the vertical direction in the plot. Apart from these deviations, we achieve an excellent agreement between theory and experiment.

By comparing the pulse amplitudes in experiment and theory we find that the strongest applied pulses with an intensity of $P_1=300$~nW correspond to a pulse area of about $2.37\pi$. Considering Eq.~\eqref{eq:n} the expectation value of the photon number in the cavity after pulse 1 is approximately 14. Taking into account the Poissonian distribution's fluctuation of the photon number being $\Delta n=\sqrt{\overline{n}}\approx 4$ we find a situation where easily 20 rungs of the JC ladder are occupied. Convergence of the numerical simulations show that even rungs up to $n=50$ have to be taken into account.

Next, we consider a similar situation, but this time the delay between pulse 1 and pulses 2, 3 is negative $\tau_{12}=-10$~ps as depicted in Fig.~\ref{fig:4}(a). Also $P_1=P_3=40$~nW and $P_2$ is scanned in the experiment from $20$~nW to $800$~nW. Simulation and measurement are plotted in Fig.~\ref{fig:4}(b) and (c), respectively. We observe similar features, as in the previous example with a positive delay. Here, the shift of the minimum lines is even stronger and it starts around $\Theta_2=1.2\pi$, i.e. at smaller pulse areas than before. We again find that these variations of the FWM dynamics agree very well between experiment and theory. Again interruptions of the first minimum line appear in the experiment especially for large pulse areas. Yet, the overall excellent agreement for these two sets of measurements yields independent verifications of the validity of the model.

\subsubsection{Limiting cases of the FWM dynamics}
To better understand the nature of the observed FWM amplitude dynamics in the previous section, it is useful to consider limiting cases of the situation. For the moment we assume that pulse 2 and 3 are so small, that the perturbative expansion of Eq.~\eqref{eq:cavity_displacement} in orders of the displacement amplitude can be used. This means that we replace the full transformation of the density matrix by the lowest order process, which does not vanish in the phase selection process from Eq.~\eqref{eq:phase_selection}. Therefore, each of the pulses 2 and 3, as it carries a phase factor of $+1$, transforms the density matrix according to
    \begin{equation}
    \rho\rightarrow \left[a^{\dagger},\rho\right]\ ,
    \end{equation}
where we neglect numerical factors. We emphasize again, that the density matrix after pulse transformation and phase selection needs neither to be hermitian nor to have unit trace. The combined action of pulse 2 and 3, which have zero relative delay, is
    \begin{equation}\label{eq:two_pulse_chi3}
    \rho\rightarrow\left[a^{\dagger},\left[a^{\dagger},\rho\right]\right]\ .
    \end{equation}
This approximation is clearly not valid for the results in Figs.~\ref{fig:3} and \ref{fig:4}, as each of the two pulses has a pulse area of $\pi/2$, which according to Eq.~\eqref{eq:n} corresponds to an average number of injected photons of $\pi^2/16>0.5$, if the cavity is in the ground state initially. However, this approximation shall just simplify the following arguments and make them more transparent.

\begin{figure}[t]
    \centering
    \includegraphics[width=0.5\linewidth]{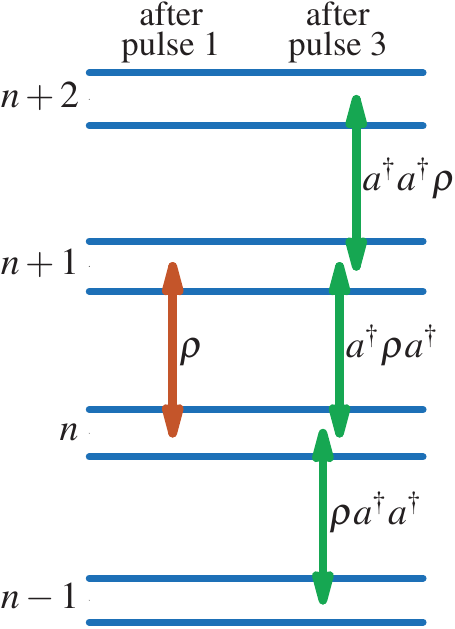}
    \caption{Illustration of the action of Eq.~\eqref{eq:two_pulse_chi3}. The left orange arrow shows a coherence after pulse 1, the right green arrows the coherences that emerge from this after pulse 3.}
    \label{fig:5}
\end{figure}

We focus on the positive delay case. The first pulse that excites the system is pulse 1, transforming the combined cavity-polaron ground state $\ket{G,0}$ into a product of a coherent cavity state and the polaron ground state $\ket{G,\alpha_1}$. The phase selection removes all parts of the density matrix, which are not one-photon coherences of the form $\ket{G,n}\bra{G,n+1}$. These one-photon coherences can be represented as superpositions of coherences between JC eigenstates of the $n$-th rung and the $(n+1)$-th rung. For the moment we neglect the influence of phonons on the polaron and any additional dephasing and decay, as described by the Lamb shift Hamiltonian and the dissipators in Eq.~\eqref{eq:Lindblad_full} and focus only on the free time evolution of the polaronic JC system in Eq.~\eqref{eq:H_S_P_main}. The density matrix entries of coherences between adjacent rungs, here $\ket{n,\sigma}\bra{n+1,\sigma'}$, evolve in time according to
    \begin{align}\label{eq:time_evol_coherences}
    &\exp\left(-\frac{i}{\hbar}H_S^Pt\right)\ket{n,\sigma}\bra{n+1,\sigma'}\exp\left(\frac{i}{\hbar}H_S^Pt\right)\\
    &=\exp\left[-\frac{i}{\hbar}(E_n^{\sigma}-E_{n+1}^{\sigma'})t\right]\ket{n,\sigma}\bra{n+1,\sigma'}\ ,\ \sigma,\sigma'=\pm\ . \notag
    \end{align}
This leads to oscillations in the photon and polaron quantities during the delay time that follows. The other two pulses transform the density matrix according to Eq.~\eqref{eq:two_pulse_chi3}, essentially mixing the old coherences between neighboring rungs into new ones as displayed in Fig.~\ref{fig:5}. The resulting density matrix now contains all coherences of the form $\ket{n+1,\sigma}\bra{n,\sigma'}$ with $\sigma,\sigma'=\pm$. The time evolution of these coherences in the density matrix is analog to Eq.~\eqref{eq:time_evol_coherences}. To obtain the final FWM signal, we have to take the trace of the density matrix together with the cavity photon annihilation operator $a$, which mixes all contributions from one-photon coherences and superimposes oscillations with the frequencies
    \begin{equation}
    \omega_n^{\sigma,\sigma'}=\frac{E_{n+1}^{\sigma}-E_n^{\sigma'}}{\hbar}=\omega_c+\frac{\sigma}{2}\widetilde{\Omega}_{n+1}-\frac{\sigma'}{2}\widetilde{\Omega}_n\ .
    \end{equation}
The cavity frequency $\omega_c$ appears in all superimposed oscillations and therefore drops out, when calculating the FWM amplitude $|\braket{a}_{\rm FWM}|$. In the resonant case with $\delta=0$, the FWM amplitude consists of superimposed oscillations with the frequencies
    \begin{equation}\label{eq:JC_frequencies}
    \left.\omega_n^{\sigma,\sigma'}\right|_{\delta=0}-\omega_c=\tilde{g}\left(\sigma\sqrt{n+1}-\sigma'\sqrt{n}\right)\ ,
    \end{equation}
which in general have different amplitudes. The ratios of all these frequencies can in general be irrational due to the special level spacing of the JC ladder. Thus, the FWM amplitude dynamics will not be periodic. Yet, we will observe dynamics that contain some kind of qualitative periodicity for sufficiently short times $t$, as irrational numbers can be approximated by rational numbers. Note, that the error of this approximation grows with $t$.

\begin{figure}[t]
    \centering
    \includegraphics[width=\linewidth]{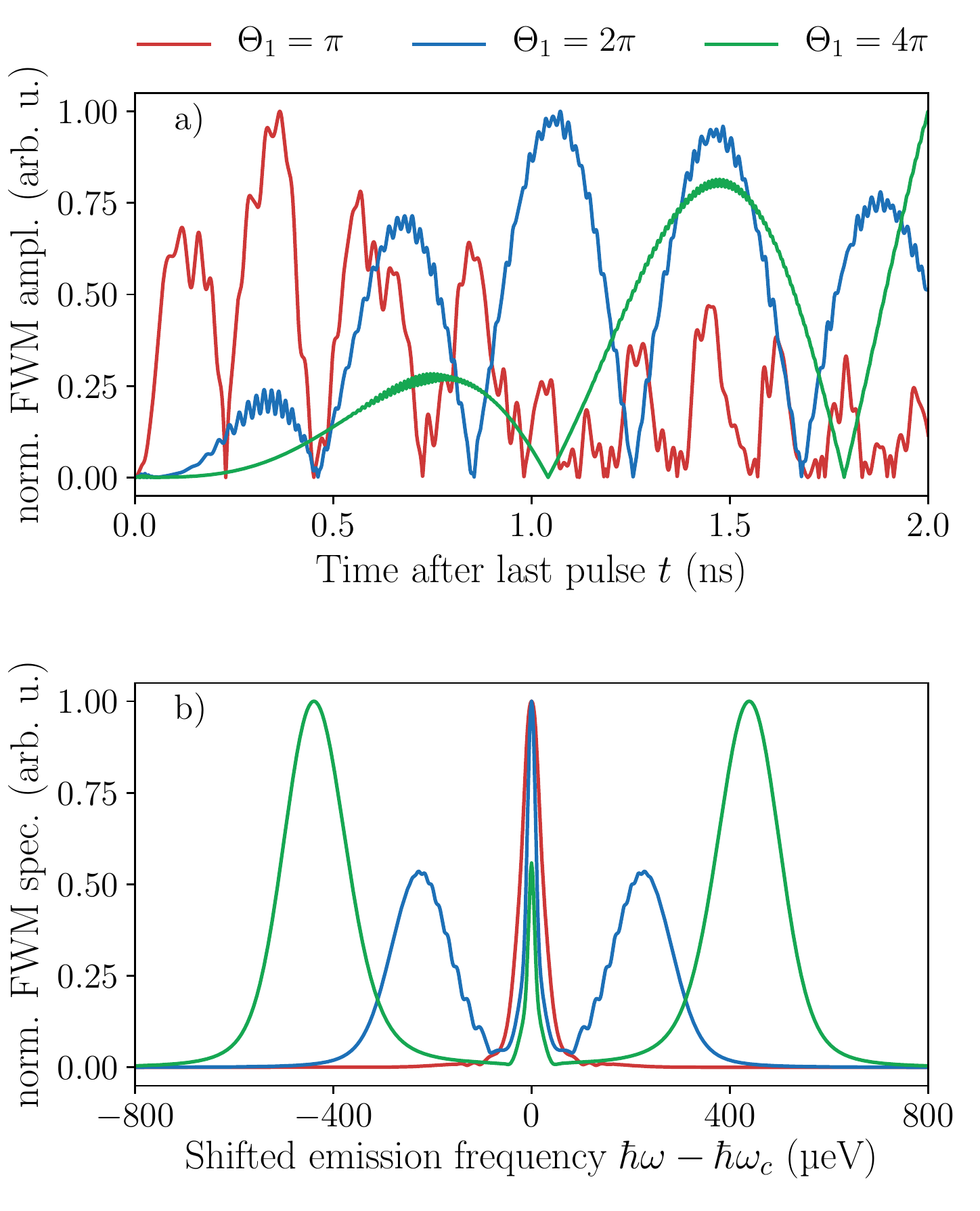}
    \caption{(a) Real time FWM amplitude dynamics for different pulse areas of pulse 1. (b) Absolute FWM emission spectrum after the excitation of the system with the last pulse for different pulse areas of pulse 1. The considered pulse areas are $\Theta_1=\pi$ (red), $\Theta_1=2\pi$ (blue) and $\Theta_1=4\pi$ (green). The calculations have been performed for the resonant case with $\delta=0$ and neglecting any dephasing and decay and renormalization due to the Lamb shift Hamiltonian. For the spectrum the spectrometer response was taken into account by multiplying the cavity field with the spectrometer response function before Fourier transforming.}
    \label{fig:6}
\end{figure}

To get an impression of the possible FWM dynamics, in Fig.~\ref{fig:6}~(a) we plot them for the three different pulse areas $\Theta_1=\pi$ (red), $2\pi$ (blue) and $4\pi$ (green) on a long timescale of 2~ns. We find that the signal evolves from an aperiodic behavior for small areas in red, that nonetheless contains some regularity, to a more and more ordered one. When $\Theta_1$ increases, two things happen to the dynamics. A dominant slow oscillation develops that is superimposed by faster components and the period of the dominant slow one increases. Because of the slow almost harmonic behavior we call these dynamics quasi-periodic. A similar superposition of oscillations leads to the well known collapse and revival phenomenon in the JC system~\cite{eberly1980periodic}. We emphasize again, that in contrast to Figs.~\ref{fig:3} and \ref{fig:4} we neglected here all dissipation present in the system. Therefore this quasi-periodic behavior of the FWM amplitude is not directly visible in the experiment, but leads to the shift of the minimum line.

We further simplify the system by considering the limiting case of small pulse area $\Theta_1$. Then we can also perform a perturbative expansion of the pulse transformation for pulse 1, yielding
    \begin{equation}
    \rho\rightarrow\left[a,\rho\right]\ .
    \end{equation}
Thus, before the arrival of the two other pulses, the density matrix is of the form
    \begin{equation}
    \rho_-=c_1 \ket{G,0}\bra{G,1}+c_2\ket{G,0}\bra{X,0}\ .
    \end{equation}
Via Eq.~\eqref{eq:two_pulse_chi3} this is transformed by the other two pulses to
    \begin{align}
    \rho_+=\sqrt{2}c_1\ket{G,2}\bra{G,1} & \notag\\
        -2c_1\ket{G,1}\bra{G,0}& \notag\\
        +\sqrt{2}c_2\ket{G,2}\bra{X,0}&\ .
    \end{align}
This density matrix contains all one-photon coherences between the ground state and the first rung and between the first and second rung. The FWM amplitude thus contains the six frequencies $\pm\tilde{g}(\sqrt{2}+1)$, $\pm\tilde{g}$, and $\pm\tilde{g}(\sqrt{2}-1)$ in the resonant case. In this case the dynamics is still simple, as not many frequencies are involved. The higher the pulse area $\Theta_1$ is, the more rungs of the JC ladder take part in the dynamics, rendering the situation hard to understand, as can be seen by inspecting the FWM amplitude dynamics in Fig.~\ref{fig:6}~(a) for the case of $\Theta_1=\pi$.

But there is another limiting case, which can be understood more easily. The average number of photons injected into the cavity by pulse 1 is $|\Theta_1|^2/4$. The uncertainty in the number of photons is given by $|\Theta_1|/2$. Thus, the relative uncertainty shrinks with the pulse area of the first pulse. Additionally, the Rabi splittings of neighboring rungs become more and more equal for large photon numbers, as $\sqrt{n+1}$ and $\sqrt{n}$ converge for large $n$. If we assume that pulse 1 injects many photons into the microcavity, such that $\sqrt{\bar{n}\pm1}\approx\sqrt{\bar{n}}$ and that we can neglect the uncertainty, all relevant rungs exhibit an equal splitting of $2\sqrt{\bar{n}}g$. Then, according to Eq.~\eqref{eq:JC_frequencies} only three frequencies are contributing to the FWM signal, namely $\omega_c$ and $\omega_c\pm\sqrt{\bar{n}}g$. Therefore the corresponding real time FWM spectrum forms a triplet, whose separation grows linearly with $|\Theta_1|$. This is a similar situation as for the famous Mollow triplet in resonance fluorescence spectroscopy~\cite{mollow1969power}.

The normalized absolute value real time FWM spectrum, obtained by Fourier transforming the FWM field $\left< a\right>_{\rm FWM}$, is displayed in Fig.~\ref{fig:6}~(b). For the calculation of the spectrum, the spectrometer response function was multiplied to the cavity field before Fourier transforming. The colors of the lines correspond to the pulse areas $\Theta_1$ in Fig.~\ref{fig:6}~(a). While we find a single peak at the cavity frequency $\omega_c$ for $\Theta_1=\pi$ (red), for larger pulse areas three peaks appear. As predicted before, one remains at $\omega_c$ and the two others move further away from the center peak with growing pulse area. The reason for taking into account the spectrometer function is, that the triplet structure in the case of large pulse areas is easier to see. When one does not take the spectrometer response into account, all spectral features are much sharper due to the infinite lifetime, considered here. Therefore all contributing coherences would show up as individual lines. Each peak would be hardly visible but by the convolution with the spectrometer the well visible triplet structure emerges.

As shown in this section, especially in Fig. \ref{fig:6} (a), the dynamics of the FWM amplitude depends strongly on the pulse area of pulse 1. Consequently, the position of the first minimum of the FWM amplitude in Figs. \ref{fig:3} and \ref{fig:4} has a non-trivial dependence on the pulse area, stemming from the superposition of a multitude of oscillations. However, subsequent lines of minima seem to have the same form as the first line. This feature is highlighted in the following section.

\subsubsection{Influence of phonon coupling and cavity dissipation}
\begin{figure}[t]
	\centering
	\includegraphics[width=\linewidth]{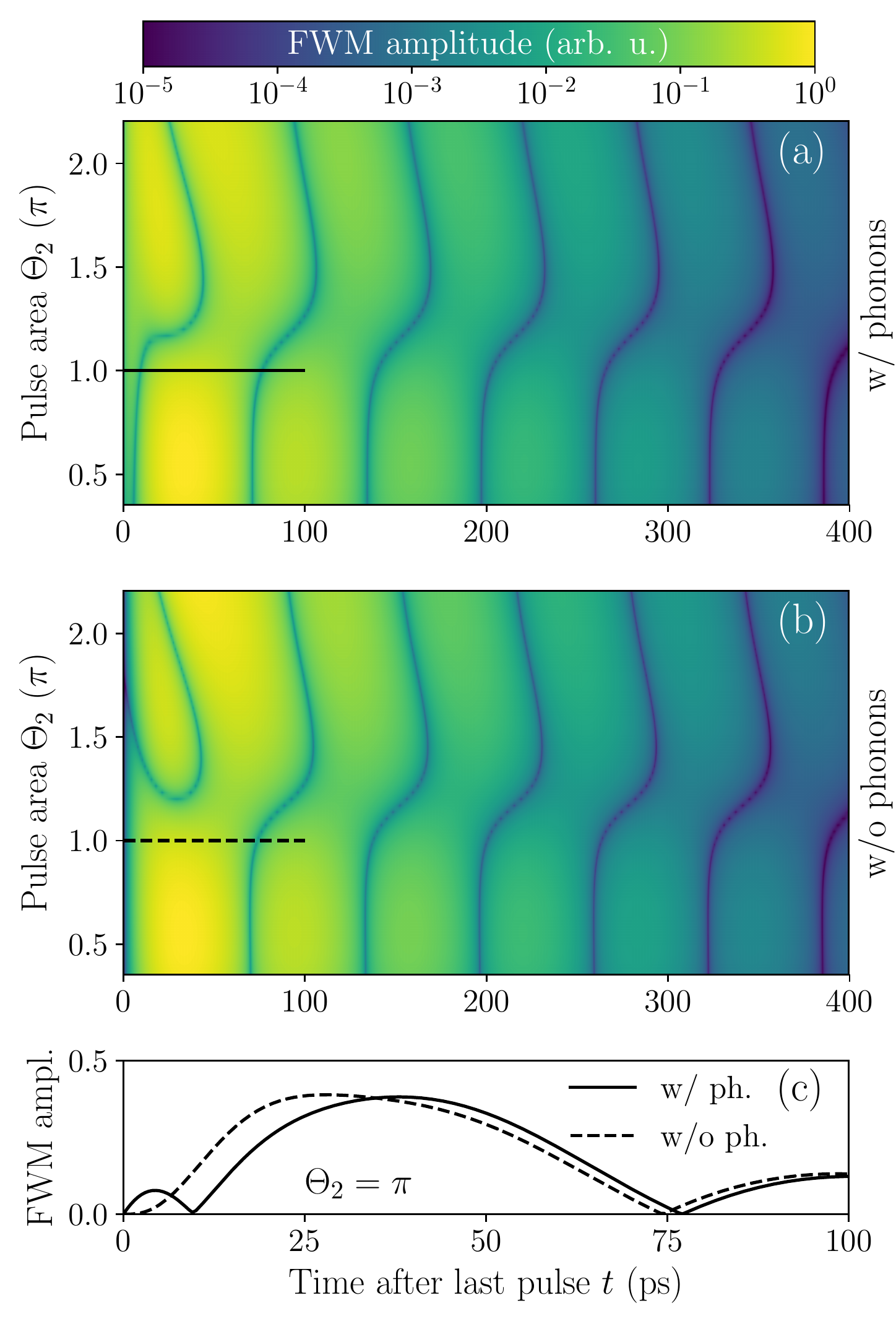}
	\caption{(a) and (b) Simulated real time dynamics of the FWM amplitude as a function of the pulse area $\Theta_2$. The pulse areas of the other two pulses are $\Theta_1=\Theta_3=\frac{\pi}{2}$. The delay between pulse 1 and pulses 2 and 3 is $\tau_{12}=-10$~ps. Pulse 2 and 3 excite the system at the same time, i.e., $\tau_{23}=0$. The situation is similar to Fig.~\ref{fig:4}, apart from the neglected spectrometer response and that no background was added to the data. Simulation presented (a) with phonon interaction and (b) without phonon interaction. (c) Comparison of the simulations with (solid line) and without (dashed line) phonons at $\Theta_2=\pi$, as marked with horizontal black lines in (a) and (b).}
	\label{fig:7}
\end{figure}
The easiest way to get an impression of the impact of the exciton-phonon coupling is to compare simulations of the full model with those disregarding the interaction, i.e. the polaron-phonon interaction as described by the dissipator $\mathcal{D}$ and the Lamb shift Hamiltonian $H_{LS}$ in Eq.~\eqref{eq:Lindblad_full}. In the following we do so and additionally neglect the spectrometer response, in contrast to Figs.~\ref{fig:3} and \ref{fig:4}, which there only led to an additional damping. Also we do not include a background in the calculations, to make the long-time behavior visible. But we again include cavity and exciton decay. With this we perform the same simulation as for Fig.~\ref{fig:4}~(b), the delay is $\tau_{12}=-10$~ps and the intensity of pulse 2, i.e. $P_2$, is scanned. The resulting FWM amplitude dynamics are shown in Fig.~\ref{fig:7}. The simulation including the interaction of the polaron with phonons is displayed in Fig.~\ref{fig:7}~(a) and the simulation neglecting the interaction with phonons is displayed in Fig.~\ref{fig:7}~(b). Overall we find agreeing features when comparing both simulations. The same modulations of the oscillations show up. However, one striking difference appears for small times $t\lesssim10$~ps. In the full simulation in Fig.~\ref{fig:7}~(a) (see also Fig.~\ref{fig:4}~(b)) an additional, almost vertical, minimum line shows up at $t\approx10$~ps. This line is absent in Fig.~\ref{fig:7}~(b), such that the dynamics starts with a minimum at $t=0$ and the next minimum appears at $t\approx 75$~ps. To have a closer look at the described dynamics within the first 100~ps, in Fig.~\ref{fig:7}~(c) we plot the FWM dynamics at $\Theta_2=\pi$, marked by the solid and dashed lines for simulations with and without phonon coupling in (a,b), respectively. Here we clearly see that the two depicted curves significantly deviate on the first 30~ps, where the phonon coupling leads to the discussed additional minimum. We conclude, that this feature at short times stems from the dissipation due to phonons. This first minimum is also clearly visible in the experiment in Fig.~\ref{fig:4}~(c). Thus we have a visible influence of the phonons on the dynamics which has to be taken into account for a sufficient description of the QD-microcavity system.

Focusing on long timescales $t>50$~ps in Fig.~\ref{fig:7} we find that the signal has the same periodicity for each pulse area $\Theta_2$. In this simulation more periods are visible than before in Figs. \ref{fig:3} and \ref{fig:4} because there the spectrometer response and background led to a rapid vanishing of the signal. Also the dynamics are calculated for longer times $t$. While in the context of Fig.~\ref{fig:6}~(a) we found that many frequencies contribute on short time scales, we can explain the survival of a single frequency for long times as follows. For the moment we neglect the influence of the exciton decay, as the cavity decay is more important for the chosen parameters. We take a look at the cavity dissipator
    \begin{equation}
    \mathcal{D}_a(\rho)=\gamma_a\left(a\rho a^{\dagger}-\frac{1}{2}\left\lbrace a^{\dagger}a,\rho\right\rbrace\right)\ .
    \end{equation}
Acting with $\mathcal{D}_a$ on an arbitrary density matrix element $\ket{m}\bra{n}$, the $\left\lbrace a^\dagger a,\rho\right\rbrace$ term generates a part proportional to $(-\ket{m}\bra{n})$ describing the decay of the $\ket{m}\bra{n}$ coherence in Eq.~\eqref{eq:Lindblad_full}. The $a\rho a^{\dagger}$ term generates a term proportional to $\ket{m-1}\bra{n-1}$, leading to an increase of this element in the density matrix. In summary the dissipator describes emission of photons from the cavity. An element $\ket{m}\bra{n}$ is converted to an element with one photon less $\ket{m-1}\bra{n-1}$, as schematically shown in Fig.~\ref{fig:2}. The larger the photon content in the density matrix components, the faster the decay, since the matrix element $\bra{n-1}a\ket{n}=\sqrt{n}$ grows with the number of photons. After the last pulse, due to phase selection, the density matrix contains all coherences between the $(n+1)$-th and the $n$-th rung. The cavity dissipation results in a cascading decay of these coherences down the JC ladder until they finally end up in the coherences between the ground state and the first rung. These coherences have the smallest photon content making them the longest-lived. The timescale of this process depends on the cavity decay rate $\gamma_a$. Thus, whatever the pulse area of pulse 2 is, for long enough times the density matrix only contains two contributions $\ket{1,\pm}\bra{G,0}$. According to Eq.~\eqref{eq:JC_frequencies}, these lead to oscillations of the FWM amplitude with the frequency $2\tilde{g}$. With the considered $\hbar\tilde{g}=35$~\textmu eV the FWM period is approximately 60~ps, which is what we find in Fig.~\ref{fig:7}. This explains the periodic behavior of the real time FWM amplitude for times $t>50$ps. Thus the influence of higher rungs of the JC ladder on the FWM amplitude dynamics is restricted to short times.

\subsection{Pulse area dependent delay dynamics}
In this section we study the traditional FWM delay dynamics~\cite{kasprzak2010up,kasprzak2013coherence} by applying the three pulse sequence from Fig.~\ref{fig:3}~(a) and scan the delay after the first pulse $\tau_{12}$, while keeping $\tau_{23}=0$ fixed. For each delay the FWM signal is retrieved by integrating over the absolute value of the spectrum and carries information on the coherence dynamics of the system.
\subsubsection{Positive delay $\tau_{12}$}
\begin{figure}[t]
    \centering
    \includegraphics[width=\linewidth]{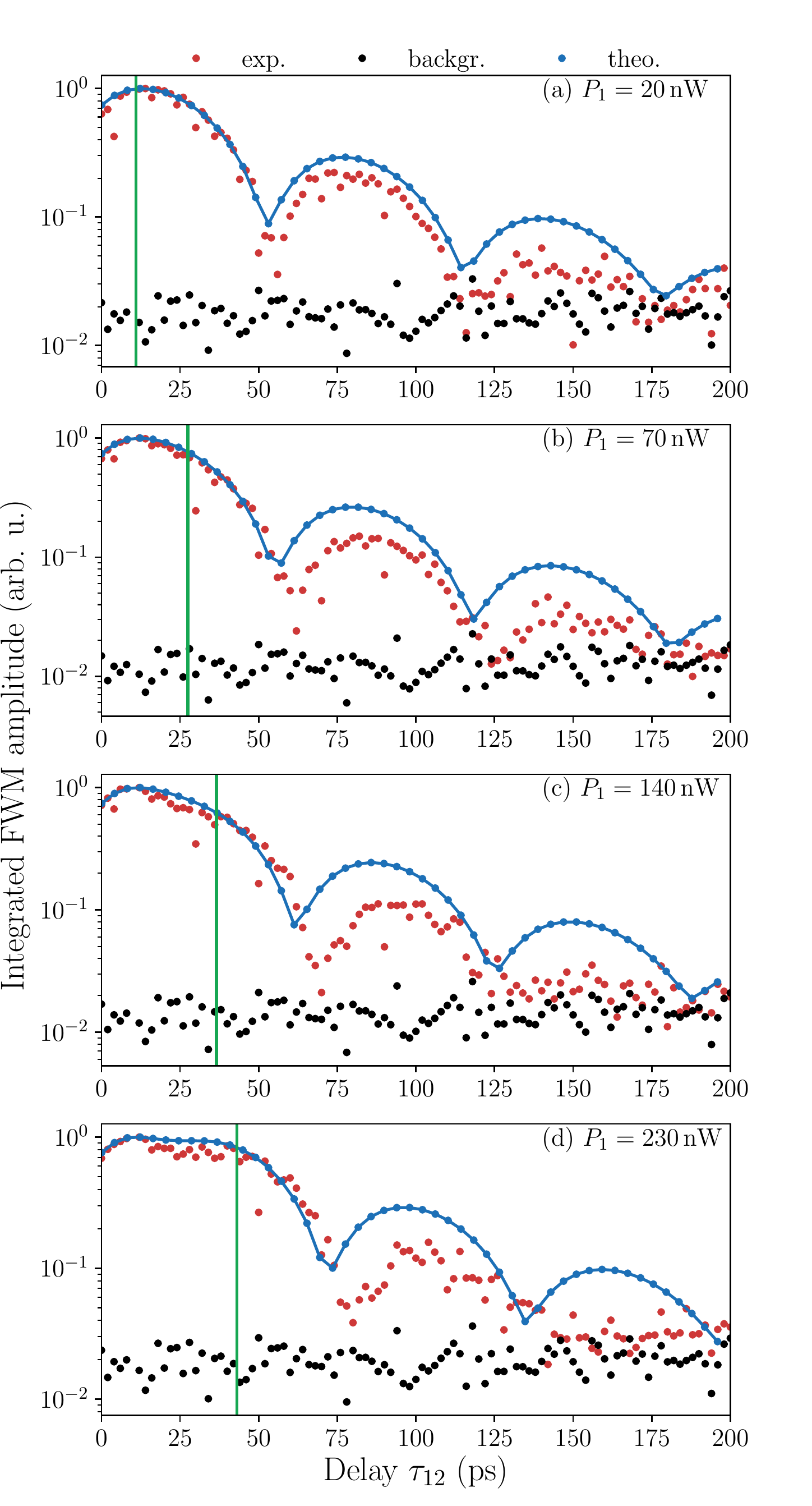}
    \caption{Delay dynamics of the integrated FWM amplitude for positive delay $\tau_{12}>0$. $P_2=P_3=40\,$nW, $P_1$ increases from top (a) to bottom (d) and is given in the figures. Measured delay dynamics (red dots) are compared to simulated delay dynamics (blue dots and line). The influence of phonons on the polaron is neglected. The background has been added to the calculations and is displayed additionally by black dots. The vertical solid green line denotes the time, when the dynamics is dominated by the first rung in a simple cavity-only model, as given by Eq.~\eqref{eq:dominance}.}
    \label{fig:8}
\end{figure}

Figure~\ref{fig:8} shows measured (red dots) and simulated (blue dots and line) FWM delay dynamics for increasing intensities of the first laser pulse from top to bottom as labeled in each plot. For the simulations we choose exactly the same parameters as for the real time dynamics in the previous section and the background (black dots) has been added to the simulation. As we found that the properties of the measured sample slightly changed from cooling cycle to cooling cycle there will be quantitative deviations between experiment and theory. However, the basic qualitative features are not affected. Here, we also neglect the interaction between the polaron and the phonons, as described by the phonon dissipator and the Lamb shift Hamiltonian in Eq.~\eqref{eq:Lindblad_full}, to keep the calculations numerically feasible. In fact we found, that for delays up to 100~ps no significant changes occur in the delay dynamics, when phonons are included, as can be seen in the appendix in Fig.~\ref{fig:12}.

For the smallest considered pulse power of $P_1=20$~nW in Fig.~\ref{fig:8}~(a), we observe damped oscillations of the measured FWM signal (red), that vanish in the background of the experiment for delays $\tau_{12}\gtrsim 175$~ps. The corresponding theoretical simulations (blue) show the same qualitative features. We observe a good agreement between simulation and experiment, as the positions of the minima agree very well. From the simulated pulse area $\Theta_1\approx0.35\pi$ we know that approximately $\Theta_1^2/4\approx 0.3$  photons are present in the cavity after the first pulse. Therefore the frequency of the oscillation is dominated by the coherences between the ground state and the first rung of the JC ladder. The corresponding period of $\pi/\tilde{g} \approx 60$~ps agrees very well with the results.

Moving to larger pulse areas of the first pulse in (b)-(d), two prominent effects can be identified:\\
(A) the first minimum shifts to later delays. While it appears at $\tau_{12}=50$~ps in (a) for $P_1=20$~nW it is at $\tau_{12}=75$~ps in (d) for $P_1=230$~nW. After this minimum the dynamics are dominated by the single frequency from the lowest JC ladder step. Therefore the shift of the first minimum and the accompanied additional dynamics around $\tau_{12}=25$~ps, which form a plateau-like structure in Fig.~\ref{fig:8}~(d), stem from contributions of higher rungs of the JC ladder. This finding is in line with the discussions in the previous section.\\
(B) For small powers of the first pulse $P_1$ in Fig.~\ref{fig:8}~(a) and (b) three periods of the oscillation are well resolved in the experiment and the noise level is reached around $\tau_{12}=175$~ps. When increasing the pulse power in Fig.~\ref{fig:8}~(c) and (d) the visibility of the signal lasts less long. This shows that the dephasing in the system works more efficient for stronger excitations. The effect is not well reproduced by the simulation because the blue curve is obviously not damped strongly enough. This could be explained by some sort of excitation induced dephasing process in the system which we did not take into account. However, it could also be the case, that the agreement between experiment and theory would be improved by another set of parameters in the large parameter space, consisting of the JC coupling $\tilde{g}$, the cavity decay rate $\gamma_a$, the exciton decay rate $\gamma_X$ and the proportionality factor $c$, relating pulse areas with laser pulse intensities.

In the following we take a closer look at effect (A). As explained before, the shift of the first minimum is due to the increased time, that the system needs to relax to a state, where the first rung dominates the coherence dynamics. To estimate this time, we consider the simplest case: We take a look at the situation, where $\tilde{g}=\gamma_X=0$, or in other words, where the dynamics of the JC system is dominated by the emission of photons from the cavity. Due to the disregarded coupling and since the laser pulses only excite the cavity mode, we restrict the dynamics in Eq.~\eqref{eq:Lindblad_full} to
    \begin{align}
        \hbar\frac{\text{d}}{\text{d}t}\rho_c(t)=&-i\left[\omega_c a^{\dagger}a,\rho_c(t)\right]+\mathcal{D}_a\left[\rho_c(t)\right]\ .
    \end{align}
$\rho_c$ denotes the density matrix of the cavity system. The time evolution of moments of the form $\braket{(a^{\dagger})^ma^m}$ is given by
    \begin{align}
        \frac{\text{d}}{\text{d}t}\braket{(a^{\dagger})^ma^m}(t)&=\text{Tr}_c\left[(a^{\dagger})^ma^m\frac{\text{d}}{\text{d}t}\rho_c(t)\right]\notag\\
        &=\hbar^{-1}\text{Tr}_c\left\lbrace(a^{\dagger})^ma^m \mathcal{D}_a\left[\rho_c(t)\right]\right\rbrace\notag\\
&=-\gamma_a m \braket{(a^{\dagger})^ma^m}(t)\ ,
    \end{align}
where Tr$_c$ denotes the trace over the cavity Hilbert space. For an initial coherent state, with amplitude $\alpha$, the mean photon number evolves as
    \begin{equation}
        \bar{n}(t)=\braket{a^{\dagger}a}(t)=e^{-\gamma_at}\bar{n}(0)=e^{-\gamma_at}|\alpha|^2\,.
    \end{equation}
The deviation $\Delta n$ from the mean number evolves as
    \begin{equation}
        \Delta n(t)=\sqrt{\braket{a^{\dagger}aa^{\dagger}a}(t)-\braket{a^{\dagger}a}^2(t)}=e^{-\frac{1}{2}\gamma_at}|\alpha|\,.
    \end{equation}
As a condition to find a dominating coherence between ground state and first rung in the dynamics, we choose
    \begin{align}\label{eq:dominance}
        \bar{n}(t)+\Delta n(t)&\leq 0.5\notag\\
        \Rightarrow t&\geq \frac{2}{\gamma_a}\ln\left(\frac{2|\alpha|}{\sqrt{3}-1}\right)\ .
    \end{align}
This condition simply states, that the coherence between ground state and first rung is dominant, when the number of photons is sufficiently small. In the case of the delay dynamics, presented in Fig.~\ref{fig:8}, the absolute value of the initial coherent amplitude is given by
    \begin{equation}
        |\alpha|=\frac{|\Theta_1|}{2}=\frac{c\sqrt{P_1}}{2}\ ,
    \end{equation}
where $c=2.5\pi\sqrt{\text{\textmu W}}^{-1}$ is the proportionality factor between pulse area and laser field strength. With our standard parameter for the cavity decay of $\hbar\gamma_a=50$~\textmu eV, we can estimate the time from which the first rung coherence dominates by this simplified model. For the considered laser intensities $P_1$, these times are marked in Fig.~\ref{fig:8} as green vertical lines. We observe, as best seen in Fig.~\ref{fig:8} (d), that the time when the first rung becomes dominant marks the end of the plateau-like structure and damped single frequency oscillations occur in the delay dynamics from that point on. Therefore, we observe that even this simple model shows the same qualitative behavior in terms of the dominance of the first rung coherences. Although it does not perfectly resemble the situation in the experiment, because the influence of the exciton is fully neglected, we can easily understand the origin of the shift of the first minima in Fig.~\ref{fig:8} from it.

\subsubsection{Negative delay $\tau_{12}$}
\begin{figure}[t]
    \centering
    \includegraphics[width=\linewidth]{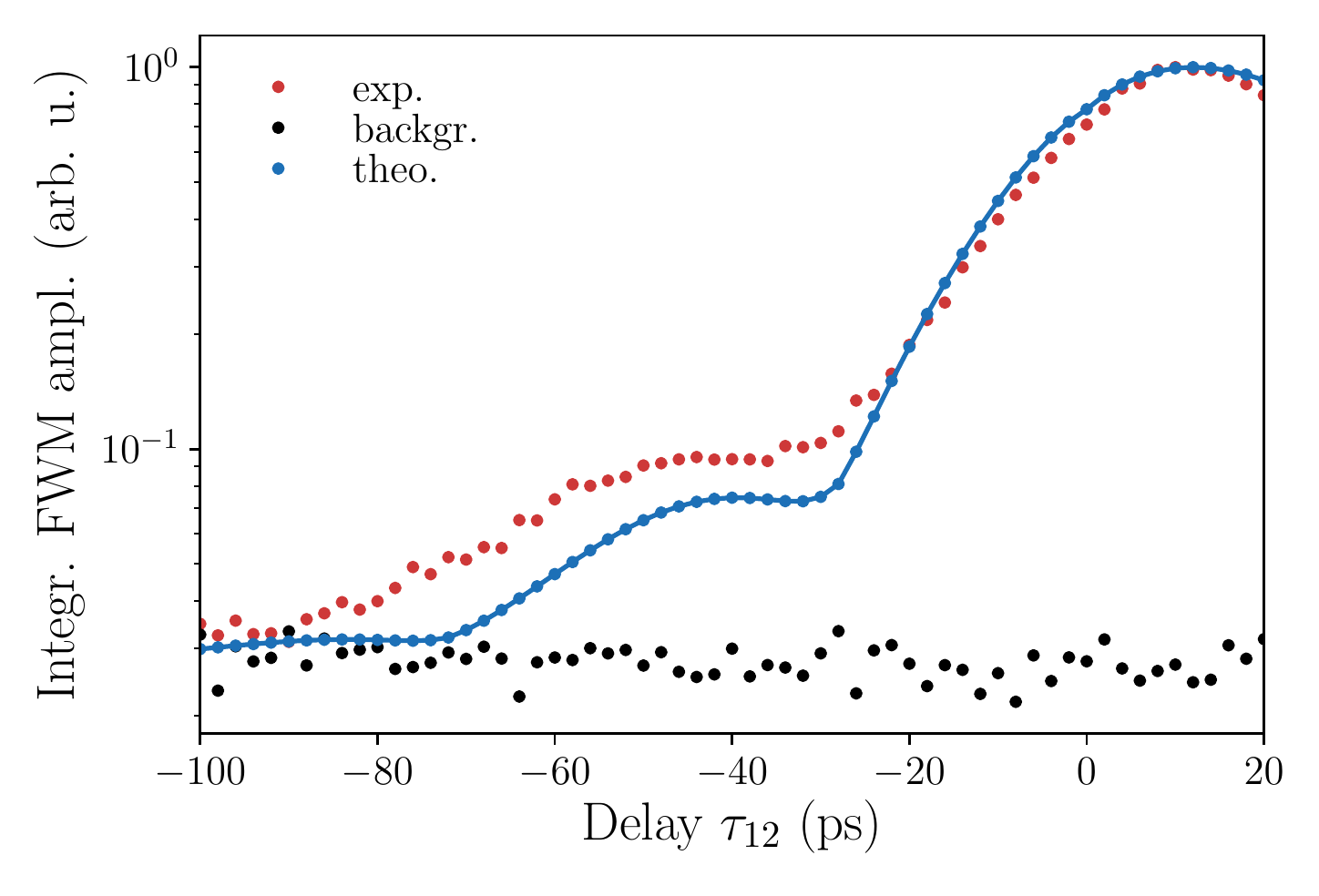}
    \caption{Delay dynamics for a negative delay $\tau_{12}$. The laser pulse intensities are $P_1=P_2=P_3=25$~nW. Measured dynamics (red dots) are compared to simulated dynamics (blue dots and line). The influence of phonons on the polaron is neglected. The background (black dots) has been added to the simulation.}
    \label{fig:9}
\end{figure}

Finally, we present the case, where the delay $\tau_{12}$ is scanned over negative values. The laser intensities are given by $P_1=P_2=P_3=25$~nW. In Fig.~\ref{fig:9} the measured delay dynamics (red dots) is compared to the simulation (blue dots and line). The background of the experiment (black dots) has been added to the simulation.

In the case of negative delay, the dynamics do not stem from coherences between neighboring rungs. In this case only two-photon coherences, which are coherences between the $n$-th and $(n+2)$-th rung contribute to the dynamics. For sufficiently low pulse areas, one observes damped oscillations, originating from the ground state to second rung coherence. In analogy to Eq.~\eqref{eq:time_evol_coherences}, the frequency of this oscillation is given by $\widetilde{\Omega}_2=2\tilde{g}\sqrt{2}$ in the case that $\delta=0$. Thus we expect oscillations with a period, that is shortened by a factor of $1/\sqrt{2}$, compared to the positive delay case. This can in fact be observed in Fig.~\ref{fig:9}. While in the positive delay case for small pulse areas in Fig.~\ref{fig:8} (a) the period of the beat is seen to be $2\pi/\widetilde{\Omega}_1\approx 60$~ps, here it is $2\pi/\widetilde{\Omega}_2\approx 42$~ps. Note, that $42/60=0.7\approx 1/\sqrt{2}$. This can be seen from the distance of the minima in the simulation. The oscillation dynamics of the two-photon coherence is not as pronounced in the experiment, but overall there is a good agreement between experiment and simulation also in the case of negative delay $\tau_{12}$.
\section{Conclusions}
By combining FWM micro-spectroscopy with detailed numerical simulations we have investigated the dynamics of a quantum dot-microcavity system in the regime where many photons, up to $\sim 20$, are injected into the cavity by strong laser pulses. The simulations of the FWM signals are in excellent agreement with the experiment, verifying the validity of the JC model coupled to LA phonons in this regime. The investigation of the real time FWM dynamics has shown that at short times after the pulse excitation the dynamics are quite involved, as many rungs of the JC ladder take part in it. For long times however, single frequency oscillations occur. The simulations showed, that in a system without dissipation, the FWM dynamics have a quasi-periodic behavior, of the same origin as the collapse and revival phenomenon in the JC system. It was predicted, that in principle the measured FWM spectrum exhibits a triplet structure similar to the Mollow triplet in resonance fluorescence. However, the visibility of this triplet structure is limited by the dissipation in the JC system. The FWM delay dynamics at positive delay showed similar features as the real time dynamics. At small delays, shortly after the first pulse excitation, the dynamics strongly depend on the number of injected photons, exhibiting a plateau-like structure, whose length grows with the number of photons. The negative delay FWM signal showed a beat with a period that is about a factor $\sqrt{2}$ shorter than the positive delay FWM signal, verifying the level structure of the JC ladder. The influence of phonons on the FWM signals was only visible in the real time dynamics, leading to an additional minimum of the signals at short times and small pulse areas. The delay dynamics was not strongly affected by the phonons, as shown by comparing simulations with and without phonons. Our results demonstrate that FWM micro-spectroscopy is a powerful technique to investigate the JC system even in a regime, where many rungs of the JC ladder are important for the dynamics. In the future, general $N$-wave mixing could be used to resolve the $\sqrt{n}$-dependence of the JC ladder in the delay dynamics for more than the first two rungs. This could be possible even in cases, where the JC ladder cannot be resolved spectrally by common linear spectroscopy techniques. Here, $N$-wave mixing allows to isolate multi-photon coherences and observe their specific dynamics.
\begin{acknowledgements}
	D.W. acknowledges financial support by the Polish National Agency for Academic Exchange (NAWA) within the ULAM program (No. PPN/ULM/2019/1/00064).	The W\"urzburg team acknowledges the support by the State of Bavaria and the Deutsche Forschungsgemeinschaft (DFG) within Project No. SCHN1376 5.1  / PR1749 1.1.
\end{acknowledgements}
\appendix
\section{JC model in the polaron frame}
\label{sec:polaronframe}
As described in Sec.~\ref{sec:model}, we transform the system, which is governed by the Hamiltonian from Eq.~\eqref{eq:H_tot}, to the polaron frame before employing the master equation approach. The unitary operator describing this transformation is given by~\cite{roy2011influence,nazir2016modelling,mahan2013many}
\begin{equation}\label{eq:polaron_transform}
\exp(S)\equiv\exp\left[X^{\dagger}X\sum_{\bf q}\frac{g_q}{\omega_q}\left(b_{\bf q}^{\dagger}-b_{\bf q}^{}\right)\right]\,.
\end{equation}
Note, that operators in the polaron frame will be denoted by a superscript $P$, i.e.
\begin{equation}
\hat{O}^P=\exp(S)\hat{O}\exp(-S)\ .
\end{equation}
It is important to notice, that photonic operators are not affected by this transformation. The Hamiltonian in Eq.~\eqref{eq:H_tot} then takes the following form in the polaron frame
\begin{subequations}
    \begin{align}
    H^P&=H_S^P+H_B^P+H_I^P\\
    H_S^P&=\hbar\omega_c a^{\dagger}a+\hbar\tilde{\omega}_X X^{\dagger}X+\hbar\tilde{g}\left(aX^{\dagger}+a^{\dagger}X\right) \label{eq:H_S_P}\\
    H_B^P&=\sum_{\bf q}\hbar \omega_q b_{\bf q}^{\dagger}b_{\bf q}^{}\\
    H_I^P&=\underbrace{a^{\dagger}X^{}_{}}_{A^P_1}\underbrace{\hbar g_{\rm JC}(B_--B)}_{B^P_1}+\underbrace{aX^{\dagger}_{}}_{A^P_2}\underbrace{\hbar g_{\rm JC}(B_+-B)}_{B^P_2}\ .\label{eq:H_I_P}
    \end{align}
\end{subequations}
The system Hamiltonian $H_S^P$ is again of a JC form, but with a renormalized coupling strength
\begin{subequations}
    \begin{equation}
    \tilde{g}=Bg_{\rm JC}
    \end{equation}
    and a polaron-shifted exciton frequency
    \begin{equation}
    \tilde{\omega}_X=\omega_X-\sum_{\bf q}\frac{g_q^2}{\omega_q}=\omega_X-\int\limits_0^{\infty}\text{d}\omega\,\frac{J(\omega)}{\omega}\,.
    \end{equation}
    $B$ is the thermal average of the phonon displacement operators in phase space $B_\pm$ at temperature $T$~\cite{roy2011influence}
    \begin{align}\label{eq:H_tot_pol_frame}
    &B=\text{Tr}_B\left(B_{\pm}\rho_B^P\right)=\exp\left\lbrace -\frac{1}{2}\int\limits_0^{\infty}\text{d}\omega\,\frac{J(\omega)}{\omega^2}\left[2n(\omega)+1\right]\right\rbrace\\
    &B_{\pm}=\exp\left[\pm\sum_{\bf q}\frac{g_q}{\omega_q}\left(b_{\bf q}^{\dagger}-b_{\bf q}^{}\right)\right]\\
    &\rho_B^P=\frac{\exp(-\beta H_B^P)}{\text{Tr}\left[\exp(-\beta H_B^P)\right]}\,,\quad\beta=\left(k_BT\right)^{-1}\\
    &n(\omega)=\left[\exp(\beta\hbar\omega)-1\right]^{-1}\,.
    \end{align}
\end{subequations}
$\text{Tr}_B$ denotes the trace over the phonon degrees of freedom. $H_I^P$ in Eq.~\eqref{eq:H_I_P} describes the interaction between the cavity photons, the polaron and the phonons, which vanishes in the limit $g_{\rm JC}\rightarrow 0$, as we then recover the independent boson model. The operators $A_i^P$ and $B_i^P$, $i=1,2$, defined in Eq.~\eqref{eq:H_I_P} will be used below when deriving the Lindblad master equation.

The spectrum of the JC Hamiltonian $H_S^P$ in Eq.~\eqref{eq:H_S_P} can be calculated analytically. Only the states $\ket{G,n}$ and $\ket{X,n-1}$ are coupled for every $n>0$, where $n$ is the number of photons present in the cavity and is called the rung number. Thus the Hamiltonian is block-diagonal in this basis, with blocks
\begin{align}
H_S^{P(n)}&=
\begin{bmatrix}
\bra{G, n}H_S^P\ket{G, n} & \bra{G, n}H_S^P\ket{X, n-1}\\
\bra{X, n-1}H_S^P\ket{G, n} & \bra{X, n-1}H_S^P\ket{X, n-1}
\end{bmatrix}\notag\\
&=
\begin{bmatrix}
n\omega_c & \tilde{g}\sqrt{n} \\
\tilde{g}\sqrt{n} & n\omega_c -\delta
\end{bmatrix}
\label{eq:nthrungJCH}\,.
\end{align}
The ground state is given by $\ket{G,0}$ and has vanishing energy. The remaining eigenenergies are given by
\begin{subequations}\label{eq:JC_energies_states}
    \begin{equation}
    E_n^{\pm}=E_n^0\pm \frac{1}{2}\hbar\widetilde{\Omega}_n\ ,
    \end{equation}
    with
    \begin{align}\label{eq:Omega_tilde}
    E_n^0&=n\hbar\omega_c-\hbar\frac{\delta}{2}\ ,
    &\quad  \delta&=\omega_c-\tilde{\omega}_X\ ,\notag\\
    \widetilde{\Omega}_n&=\sqrt{\delta^2+\Omega_n^2}\ ,
    &\quad \Omega_n&=2\tilde{g}\sqrt{n}\ .
    \end{align}
    $\delta$ is the detuning between cavity and polaron-shifted exciton frequency, $\widetilde{\Omega}_n$ is the Rabi frequency, which coincides with the resonant Rabi frequency $\Omega_n$ in the case of vanishing detuning. This resonant Rabi frequency is not to be confused with the radio-frequencies of the acousto optical modulators, described in Sec.~\ref{sec:experiment}. Each block in Eq. \eqref{eq:nthrungJCH} is mathematically equivalent to a two-level system that is coupled to a cw-laser, detuned from the two-level systems transition energy by the detuning $\delta$~\cite{nazir2016modelling}. The Rabi frequencies of the driven two-level systems are given by $\widetilde{\Omega}_n$. Each of the two-level systems has its own ground state energy, given by $n\omega_c$.

    The eigenstates of the JC system will be needed later in the context of the master equation approach. They are given by
    \begin{equation}
    \ket{n,\pm}=\kappa_n^{\pm}\ket{G,n}+\lambda_n^{\pm}\ket{X,n-1}
    \end{equation}
    with the amplitudes
    \begin{align}
    &\kappa_n^{\pm}=\frac{\frac{\delta}{2}\pm\frac{1}{2}\widetilde{\Omega}_n}{\sqrt{n\tilde{g}^2+\left(\frac{\delta}{2}\pm\frac{1}{2}\widetilde{\Omega}_n\right)^2}}\notag\\
    &\lambda_n^{\pm}=\frac{\tilde{g}\sqrt{n}}{\sqrt{n\tilde{g}^2+\left(\frac{\delta}{2}\pm\frac{1}{2}\widetilde{\Omega}_n\right)^2}}\ .
    \end{align}
\end{subequations}

\section{Master equation approach for the exciton-phonon interaction}
We model the interaction of the QD exciton with phonons using a Lindblad master equation. Starting point is a time-convolutionless master equation of second order in perturbation theory (TCL2). Following Ref.~\cite{breuer2002theory}, we quickly review the derivation of the Lindblad master equation, in order to introduce the notation consistently and to discuss the meaning of the applied approximations.

We consider the following situation. The entire quantum system contains a part, the open system $S$, which is coupled to the remaining part, the bath $B$. The free dynamics of the open system is governed by the system Hamiltonian $H_S$ and the free dynamics of the bath by the bath Hamiltonian $H_B$. Both parts of the entire system are coupled by the interaction Hamiltonian $H_I$. Expectation values of operators, which are only defined on the open system's Hilbert space can be calculated with the reduced density matrix $\rho_S=\text{Tr}_B(\rho)$ alone. Here, $\rho$ is the density matrix of the entire system and $\text{Tr}_B$ denotes the trace over the bath's degrees of freedom. Thus if we are only interested in averages of observables of the open quantum system $S$, it is sufficient to calculate the dynamics of the reduced density matrix $\rho_S$. The equations of motion for the reduced density matrix can be derived from the full dynamics of the density matrix $\rho$, as described by the von Neumann equation with the Hamiltonian $H_S+H_B+H_I$. For the derivation of the effective equations of motion for the reduced density matrix it is convenient to work in the interaction picture defined by $H_S+H_B$. Interaction picture quantities will be denoted by a superscript $I$. The TCL2 equation -- of second order in the coupling $H_I$ -- is given by~\cite{breuer2002theory}
\begin{equation}\label{eq:TCL2}
\hbar^2\frac{\text{d}}{\text{d}t}\rho_S^I(t)=-\int\limits_{t_0}^t\text{d}\tau\,\text{Tr}_B\left[H_I^I(t),\left[H_I^I(\tau),\rho_S^I(t)\otimes\rho_B\right]\right]\ .
\end{equation}
Here, $t_0$ is the reference time, where interaction and Schr\"odinger picture coincide. Furthermore at time $t_0$ one has to assume a factorization of the total density matrix into
\begin{equation}\label{eq:rho_factor}
\rho(t_0)=\rho_S(t_0)\otimes\rho_B
\end{equation}
for the TCL2 equation to be valid. Another assumption made in its derivation is that the average of the interaction Hamiltonian taken with respect to the initial bath state $\rho_B$ vanishes, i.e. $\text{Tr}_B\left[H_I^I(t)\rho_B\right]=0$. We assume that $\rho_B$ is in a thermal state with a given temperature $T$
\begin{equation}
\rho_B=\frac{\exp(-\beta H_B)}{\text{Tr}\left[\exp(-\beta H_B)\right]}\ , 
\end{equation}
which commutes with the bath Hamiltonian. Then this condition can always be satisfied by redefining system and interaction Hamiltonian via
\begin{subequations}
    \begin{align}
    H_S&\rightarrow H_S+\text{Tr}_B(H_I\rho_B)\ ,\\
    H_I&\rightarrow H_I-\text{Tr}_B(H_I\rho_B)\ .
    \end{align}
\end{subequations}
This redefinition does not need to be performed in our case, as Eq.~\eqref{eq:H_I_P} is already in a form guaranteeing the vanishing of the interaction Hamiltonian with respect to the initial thermal bath state.

The interaction Hamiltonian can be written as a sum of tensor products of operators $A_{\alpha}$ acting on the open system and those acting on the bath $B_{\alpha}$, see Eq.~\eqref{eq:H_I_P}
\begin{equation}
H_I=\sum_{\alpha} A_{\alpha}\otimes B_{\alpha}\ .
\end{equation}
We choose the $A_{\alpha}$ to be dimensionless, such that the $B_{\alpha}$ have the dimension of energy. It is convenient to introduce the so-called energy eigenoperators of the open system. For an operator, acting only on the open system, they are defined by
\begin{equation}
A_{\alpha}(\hbar\omega)=\sum_{\epsilon'}\sum_{\epsilon}\Pi(\epsilon)A_{\alpha} \Pi(\epsilon')\delta_{\hbar\omega,\epsilon'-\epsilon}\ .
\end{equation}
$\epsilon$ and $\epsilon'$ run over the complete spectrum of the system Hamiltonian $H_S$ and $\Pi(\epsilon)$ is the projection operator onto the eigenspace of $H_S$ with eigenvalue $\epsilon$. $A_{\alpha}(\hbar\omega)$ describes transitions between eigenstates of $H_S$, that have an energy difference $\hbar\omega$, due to the action of the operator $A_{\alpha}$. These operators have some important properties, one being
\begin{equation}
\sum_{\omega}A_{\alpha}(\hbar\omega)=A_{\alpha}\ .
\end{equation}
They also have a simple interaction picture representation, which is ultimately the reason for introducing them. The interaction Hamiltonian in the interaction picture can be represented in the following way
\begin{equation}
H_I^I(t)=\sum_{\alpha,\omega}e^{-i\omega(t-t_0)}A_{\alpha}(\hbar\omega)B_{\alpha}^I(t)\,.
\end{equation}
We now insert this representation into the TCL2 equation and define the following bath correlation functions
\begin{equation}\label{eq:bath_corr_funcs}
G_{\alpha\beta}(\tau)=\frac{1}{\hbar^2}\text{Tr}_B\left\lbrace\left[B_{\alpha}^I(\tau)\right]^{\dagger}B_{\beta}^I(0)\rho_B\right\rbrace\ .
\end{equation}
Their one-sided finite time Fourier transforms are
\begin{equation}\label{eq:gamma}
\Gamma_{\alpha\beta}(\omega,s)=\int\limits_0^s\text{d}\tau\,e^{i\omega\tau}G_{\alpha\beta}(\tau)\ .
\end{equation}
The TCL2 equation then takes the form
\begin{align}\label{eq:TCL2_energybasis}
\frac{\text{d}}{\text{d}t}&\rho_S^I(t)=-\sum_{\omega,\omega'}\sum_{\alpha\beta}e^{i(\omega'-\omega)(t-t_0)}\Gamma_{\alpha\beta}(\omega,t-t_0)\times \\
&\times\left[A_{\alpha}^{\dagger}(\omega')A_{\beta}(\omega)\rho_S^I(t)-A_{\beta}(\omega)\rho_S^I(t)A_{\alpha}^{\dagger}(\omega')\right]+h.c. \notag
\end{align}
If the time scale $\tau_{\rho}$, on which $\rho_S^I$ varies, i.e. the time scale induced by the interaction $H_I$, is much longer than the oscillation time scale of the exponential factor in Eq.~\eqref{eq:TCL2_energybasis}, given by
\begin{equation}\label{eq:secular}
\tau_S\sim \text{max}|\omega-\omega'|^{-1}\,,\quad \omega\neq \omega'\,,
\end{equation}
we can perform a secular approximation and only keep the terms with $\omega=\omega'$. The validity of this approximation in the context of the JC system coupled to phonons will be discussed later.

Next we perform an additional Markov approximation. If the bath correlation functions in Eq.~\eqref{eq:bath_corr_funcs} decay on a time scale much shorter than $\tau_{\rho}$, we can choose $s\rightarrow\infty$ for the integral in Eq.~\eqref{eq:gamma}. This will lead to errors for small $s$, i.e. on short time scales compared to $\tau_\rho$. This is equivalent to choosing $t_0\rightarrow -\infty$ in Eq.~\eqref{eq:TCL2_energybasis}. For sufficiently large $\tau_{\rho}$ this error will be small, it corresponds to a coarse-graining of the combined dynamics of system and bath. In the case of a single QD coupled to LA phonons, this approximation is strongly violated, as such a system exhibits non-Markovian dynamics~\cite{breuer2016colloquium}. As we will see, in the case of a QD inside a cavity considered here, it is a valid approximation giving reasonable results in the parameter range considered, due to the adiabatic driving of the exciton by the cavity field. The final Lindblad master equation in the Schr\"odinger picture reads
\begin{equation}
\hbar \frac{\text{d}}{\text{d}t}\rho_S(t)=-i\left[H_S+H_{LS},\rho_S(t)\right]+\mathcal{D}\left[\rho_S(t)\right]\,.\label{eq:referenceBornMarkov}
\end{equation}
$H_{LS}$ is a hermitian Lamb shift operator, commuting with the system Hamiltonian $H_S$. It describes energy renormalizations of the eigenstates of $H_S$, induced by the bath. It is defined by
\begin{equation}\label{eq:H_LS}
H_{LS}=\hbar\sum_{\omega}\sum_{\alpha\beta}S_{\alpha\beta}(\omega)A_{\alpha}^{\dagger}(\omega)A_{\beta}(\omega)
\end{equation}
with
\begin{equation}
S_{\alpha\beta}(\omega)=\frac{1}{2i}\left[\Gamma_{\alpha\beta}(\omega,\infty)-\Gamma_{\beta\alpha}^*(\omega,\infty)\right]\ .
\end{equation}
$\mathcal{D}$ is a superoperator, called the dissipator, describing the non-unitary dynamics induced in the open system by the interaction with the bath. Dephasing and decay processes are included in the dissipator. It is defined by
\begin{align}\label{eq:lindblad_D}
\mathcal{D}(\rho)&=\hbar\sum_{\omega}\sum_{\alpha\beta}\gamma_{\alpha\beta}(\omega)\Big[A_{\beta}(\omega)\rho A_{\alpha}^{\dagger}(\omega) \notag\\
&\qquad\qquad\qquad\qquad -\frac{1}{2}\left\lbrace A_{\alpha}^{\dagger}(\omega)A_{\beta}(\omega),\rho\right\rbrace\Big]\ ,
\end{align}
where $\gamma_{\alpha\beta}(\omega)$ are the transition rates between eigenstates of $H_S$ with energy difference $\hbar\omega$, defined by
\begin{equation}
\gamma_{\alpha\beta}(\omega)=\Gamma_{\alpha\beta}(\omega,\infty)+\Gamma_{\beta\alpha}^*(\omega,\infty)\ .
\end{equation}

\section{Lindblad equation in the polaron frame}
To construct a Lindblad equation for the exciton-phonon interaction, we need the energy eigenoperators and the bath correlation functions from Eq.~\eqref{eq:bath_corr_funcs}. The required separation of the interaction Hamiltonian $H_I^P=A_1^PB_1^P+A_2^PB_2^P$ is described in Eq.~\eqref{eq:H_I_P}. The energy eigenoperator corresponding to $A_1^P$ can be calculated using the spectrum and the eigenstates of the JC Hamiltonian $H_S^P$ from Eqs.~\eqref{eq:JC_energies_states} and is given by
\begin{subequations}\label{eq:A_P}
    \begin{align}
    A_1^P(\hbar\omega)&= \sum_{n=1}^{\infty}\sqrt{n}\delta_{\omega,0}\big(\ket{n,+}\bra{n,+}\lambda_n^+\kappa_n^+ \notag\\
    &\qquad\qquad\qquad\qquad +\ket{n,-}\bra{n,-}\lambda_n^-\kappa_n^-\big)\notag\\
    &+\sum_{n=1}^{\infty}\sqrt{n}\delta_{\omega,\widetilde{\Omega}_n}\left(\ket{n,-}\bra{n,+}\lambda_n^+\kappa_n^-\right)\notag\\
    &+\sum_{n=1}^{\infty}\sqrt{n}\delta_{\omega,-\widetilde{\Omega}_n}\left(\ket{n,+}\bra{n,-}\lambda_n^-\kappa_n^+\right)\ .\label{eq:A_1_P}
    \end{align}
    The energy eigenoperator corresponding to $A_2^P$ is given by
    \begin{align}
    A_2^P(\hbar\omega)&= \sum_{n=1}^{\infty}\sqrt{n}\delta_{\omega,0}\big(\ket{n,+}\bra{n,+}\lambda_n^+\kappa_n^+ \notag\\
    &\qquad\qquad\qquad\qquad +\ket{n,-}\bra{n,-}\lambda_n^-\kappa_n^-\big)\notag\\
    &+\sum_{n=1}^{\infty}\sqrt{n}\delta_{\omega,\widetilde{\Omega}_n}\left(\ket{n,-}\bra{n,+}\lambda_n^-\kappa_n^+\right)\notag\\
    &+\sum_{n=1}^{\infty}\sqrt{n}\delta_{\omega,-\widetilde{\Omega}_n}\left(\ket{n,+}\bra{n,-}\lambda_n^+\kappa_n^-\right)\ .\label{eq:A_2_P}
    \end{align}
\end{subequations}
The respective second and third terms describe phonon assisted transitions within one rung as depicted in Fig.~\ref{fig:2} by the red arrow. The bath correlation functions in the polaron frame read
\begin{subequations}\label{eq:G}
    \begin{align}\label{eq:G11}
    G_{11}^P(\tau)&=G_{22}^P(\tau)=\tilde{g}^2\left\lbrace \exp\left[\phi(\tau)\right]-1\right\rbrace\\
    G_{12}^P(\tau)&=G_{21}^P(\tau)=\tilde{g}^2\left\lbrace \exp\left[-\phi(\tau)\right]-1\right\rbrace\ ,\label{eq:G12}
    \end{align}
\end{subequations}
where the function $\phi(\tau)$ is well known in the context of phonon-induced dephasing of two-level systems~\cite{wigger2019phonon, franke2019quantization}, and is given by
\begin{equation}
\phi(\tau)=\int\limits_0^{\infty}\text{d}\omega\,\frac{J(\omega)}{\omega^2}\left\lbrace[n(\omega)+1]e^{-i\omega\tau}+n(\omega)e^{i\omega\tau}\right\rbrace\ .
\end{equation}
The decay rates in the Lindblad dissipator will be calculated numerically from the bath correlation functions, which decay on a ps time-scale. Figure~\ref{fig:10} shows the phonon spectral density $J(\omega)$ for the parameters chosen in the simulations.

We account for cavity losses and (radiative) decay of the polaron (into non-cavity modes) with two phenomenological Lindblad dissipators of the form~\cite{kasprzak2010up,roy2011influence,kasprzak2013coherence, carmele2019non}
\begin{equation}\label{eq:dissipator}
\mathcal{D}_L(\rho)=\hbar\gamma_L\left(L\rho L^{\dagger}-\frac{1}{2}\left\lbrace L^{\dagger}L,\rho\right\rbrace\right)
\end{equation}
with $L=a,\,X$, respectively. The photon loss is indicated in Fig.~\ref{fig:2} by the green arrow.

Finally, the equation of motion for the reduced density matrix in the polaron frame reads
\begin{align}\label{eq:Lindblad_full_appendix}
\hbar\frac{\text{d}}{\text{d}t}\rho_S^P(t)=&-i\left[H_S^P+H_{LS},\rho_S^P(t)\right]+\mathcal{D}\left[\rho_S^P(t)\right]\notag\\
&+\mathcal{D}_a\left[\rho_S^P(t)\right]+\mathcal{D}_X\left[\rho_S^P(t)\right]\ .
\end{align}
The phonon dissipator $\mathcal{D}$ is given by Eq.~\eqref{eq:lindblad_D}, using the energy eigenoperators of $A_1^P$ and $A_2^P$ in Eqs.~\eqref{eq:A_P} and the decay rates calculated from the bath correlation functions $G_{\alpha\beta}^P$ in Eqs.~\eqref{eq:G}. Similarly, the Lamb shift Hamiltonian $H_{LS}$ is given by Eq.~\eqref{eq:H_LS}. The equation of motion from Eq.~\eqref{eq:Lindblad_full_appendix} will be used to calculate the dynamics of the system between the delta-pulses, that excite the cavity.

\section{Influence of the applied approximations}
\begin{figure}[t]
	\centering
	\includegraphics[width=\linewidth]{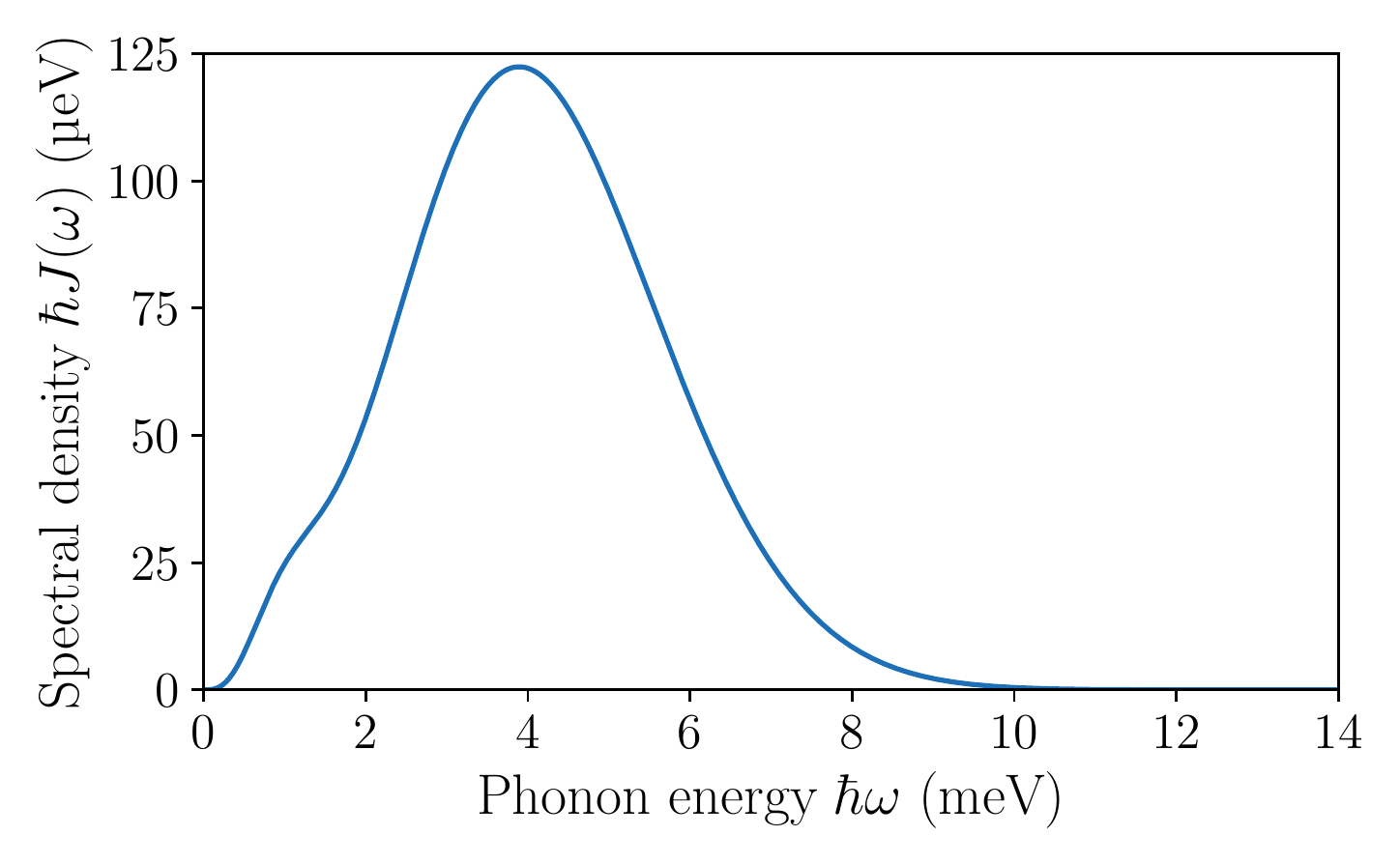}
	\caption{Phonon spectral density, as given in Eq.~\eqref{eq:specdens}.}
	\label{fig:10}
\end{figure}
    \begin{figure}[t]
    	\centering
    	\includegraphics[width=\linewidth]{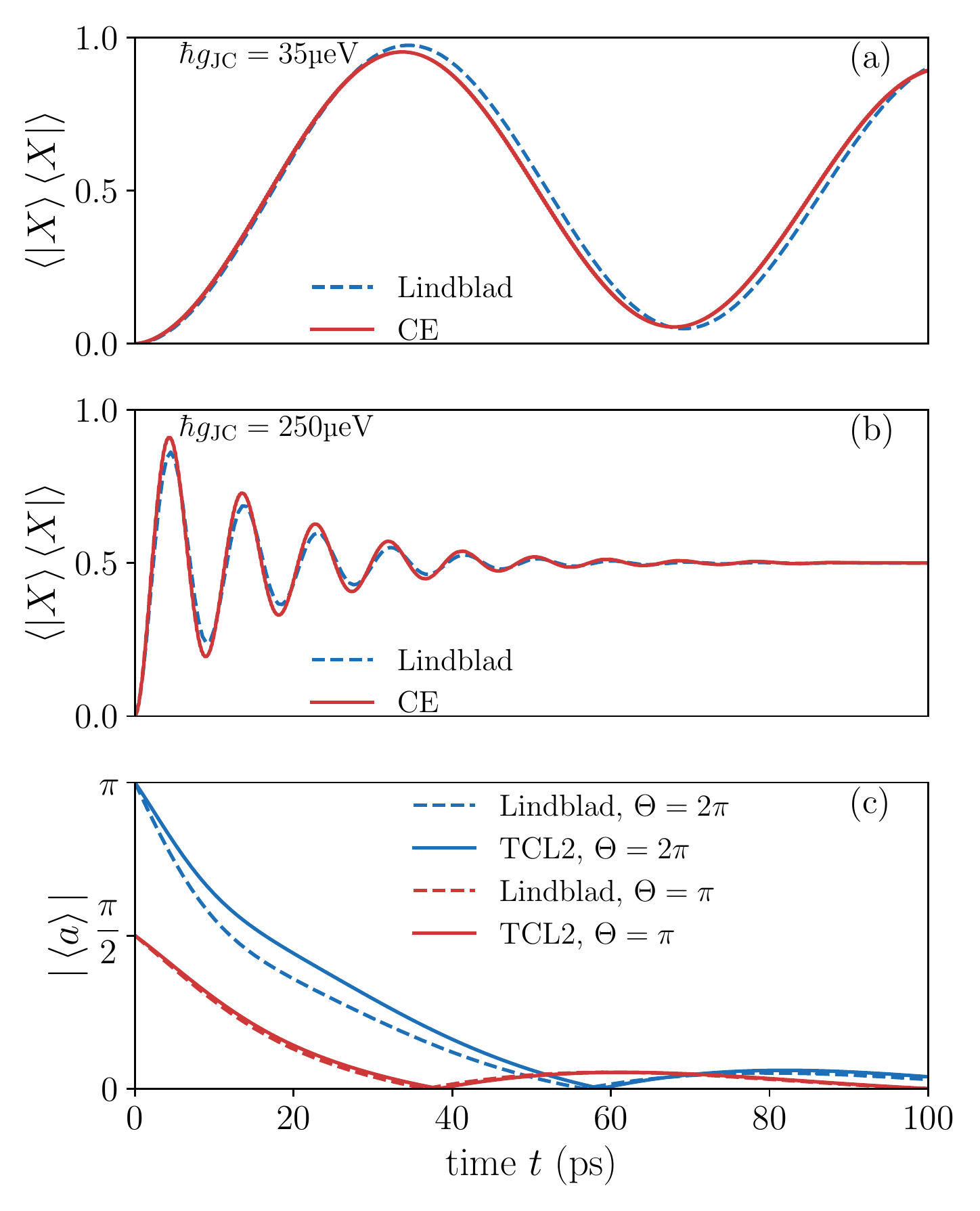}
    	\caption{(a) and (b) Comparison of vacuum Rabi oscillations in the JC system (blue) with correlation expansion (CE) calculations, including two-fold phonon-assisted quantities, of a QD driven by a cw-laser (red). Displayed is the time-dependent polaron occupation $\braket{\ket{X}\bra{X}}$. The cw-laser, as well as the cavity frequency are each resonant to the polaron frequency. The bare Rabi frequencies $2g_{\rm JC}$ (before renormalization by phonon interaction) are given in the picture. The temperature is chosen to $T=23\,K$. Cavity losses and exciton decay are not considered, $\gamma_a=\gamma_X=0$. (c) Dynamics of the absolute value of the cavity field $|\!\braket{a}\!|$, after excitation of the cavity-polaron system from its ground state by a single pulse with pulse areas $\Theta=2\pi$ (blue) and $\Theta=\pi$ (red). The calculations have been performed, using the full TCL2 Eq.~\eqref{eq:TCL2} (solid) and the Lindblad Eq.~\eqref{eq:Lindblad_full_appendix} (dashed). The temperature of the phonon bath has been chosen to $T=23$~K, as in the experiment. Likewise, the cavity and polaron decay rates have been set to $\hbar\gamma_a=50$~\textmu eV and $\hbar\gamma_X=2$~\textmu eV. The JC model parameters are given by $\hbar\tilde{g}=35$~\textmu eV and $\delta=0$.}
    	\label{fig:11}
    \end{figure}
    In the following we briefly discuss the validity of the approximations, that were used to arrive at the Lindblad equation for our specific system, as given in Eq.~\eqref{eq:Lindblad_full_appendix}. The JC system is formally identical to a collection of cw-lasers interacting with the exciton/polaron, as seen in Eq.~\eqref{eq:nthrungJCH}. The efficiency of the coupling of phonons to the polaritons of the JC system is described by the spectral density, displayed in Fig.~\ref{fig:10}. For not too large coupling strengths $\tilde{g}$ and detunings $\delta$, the Rabi frequencies of the relevant JC ladder rungs $\widetilde{\Omega}_n$ in Eq.~\eqref{eq:Omega_tilde} are small compared to the frequencies of the phonon spectral density in Eq.~\eqref{eq:specdens}, at which the coupling to phonons is efficient, in our case above 0.5~meV. In this case the polaron is adiabatically switched on and off by the collection of cw-lasers~\cite{machnikowski2007quantum,wigger2014energy} and we can assume the factorization of the density matrix in Eq.~\eqref{eq:rho_factor}. Furthermore this adiabatic driving of the two-level system prevents non-Markovian dynamics, which would be present in the case of direct pulsed excitations of the QD~\cite{krummheuer2002the}. This is the justification for the Born-Markov approximations, which have been discussed shortly before Eq.~\eqref{eq:referenceBornMarkov}.

    In Fig. \ref{fig:11}(a) and (b) we compare the dynamics of vacuum Rabi oscillations of the polaron occupation, calculated with the Lindblad approach, with correlation expansion (CE) calculations, including two-fold phonon-assisted quantities (also called fourth Born approximation), of a cw-driven QD, coupled to LA phonons, e.g. in Ref.~\cite{rossi2002theory,kruegel2005role,glaessl2011long,lengers2020TMDC}. The cavity frequency, as well as the frequency of the cw-laser in the CE calculations are chosen resonant to the polaron shifted exciton frequency $\tilde{\omega}_X$. We neglect cavity and exciton dissipation, i.e. $\gamma_a=\gamma_X=0$ and choose the temperature for the LA phonon bath to $T=23$~K, as this resembles the experimental situation. We consider two different bare Rabi frequencies $2g_{\rm JC}$ to check the validity of the approximation over a range of Rabi splittings, i.e. for a large number of JC ladder rungs for a chosen coupling $g_{\rm JC}$. The bare JC coupling $\hbar g_{\rm JC}=35$~\textmu eV in (a) is close to the coupling that fits well with the experimental data. As the mathematical structure of the JC system is the same in every rung, apart from different Rabi frequencies, we can understand the Lindblad calculation in (b) at $\hbar g_{\rm JC}=250$~\textmu eV as oscillations in an $n$-th rung, instead of vacuum Rabi oscillations. The rung number is $(250/35)^2\approx 51$. We see, that the Markovian Lindblad equation [Eq.~\eqref{eq:Lindblad_full}] can reproduce the non-Markovian correlation expansion calculations very well for the first $\approx 50$ rungs of the JC ladder. As the dominating decay and dephasing process is the spontaneous emission of photons from the cavity for the parameters relevant in this paper, we consider the small deviations of the Lindblad from the CE calculations to be of minor importance compared to the huge numerical advantage of the Lindblad method.

    The secular approximation, discussed after Eq.~\eqref{eq:secular}, is a bigger issue. It relies on the smallness of $\tau_S\sim\max|\omega-\omega'|^{-1}$ for $\omega\neq\omega'$ compared to the systems time scale $\tau_{\rho}$, where $\omega$ and $\omega'$ are energy differences in the part of the spectrum of $H_S^P$, which is relevant for the dynamics. This depends on the number of relevant JC ladder rungs, i.e. the number of photons present in the cavity. The higher the pulse areas of the considered laser pulses, the more photons will be present in the cavity. In the case of vanishing detuning, the time scale $\tau_S$ scales as
    \begin{equation}
    \tau_S\sim \left[2\tilde{g}\left(\sqrt{N}-\sqrt{N-1}\right)\right]^{-1}\,,
    \end{equation}
    where $N$ is the maximum number of photons, which are relevant for the dynamics of the system. The larger $N$ gets, the larger will the time scale $\tau_S$ be. Thus, we expect the secular approximation to break down for a large number of photons present in the cavity. On the other hand it is exactly this case, in which the secular approximation gives huge numerical advantages, as it allows to reduce the double sum in Eq.~\eqref{eq:TCL2_energybasis} to a single sum. The number of terms of the double sum scales approximately with $N^2$, whereas the number of terms of the single sum scales approximately with $N$.

\begin{figure}[t]
	\centering
	\includegraphics[width=\linewidth]{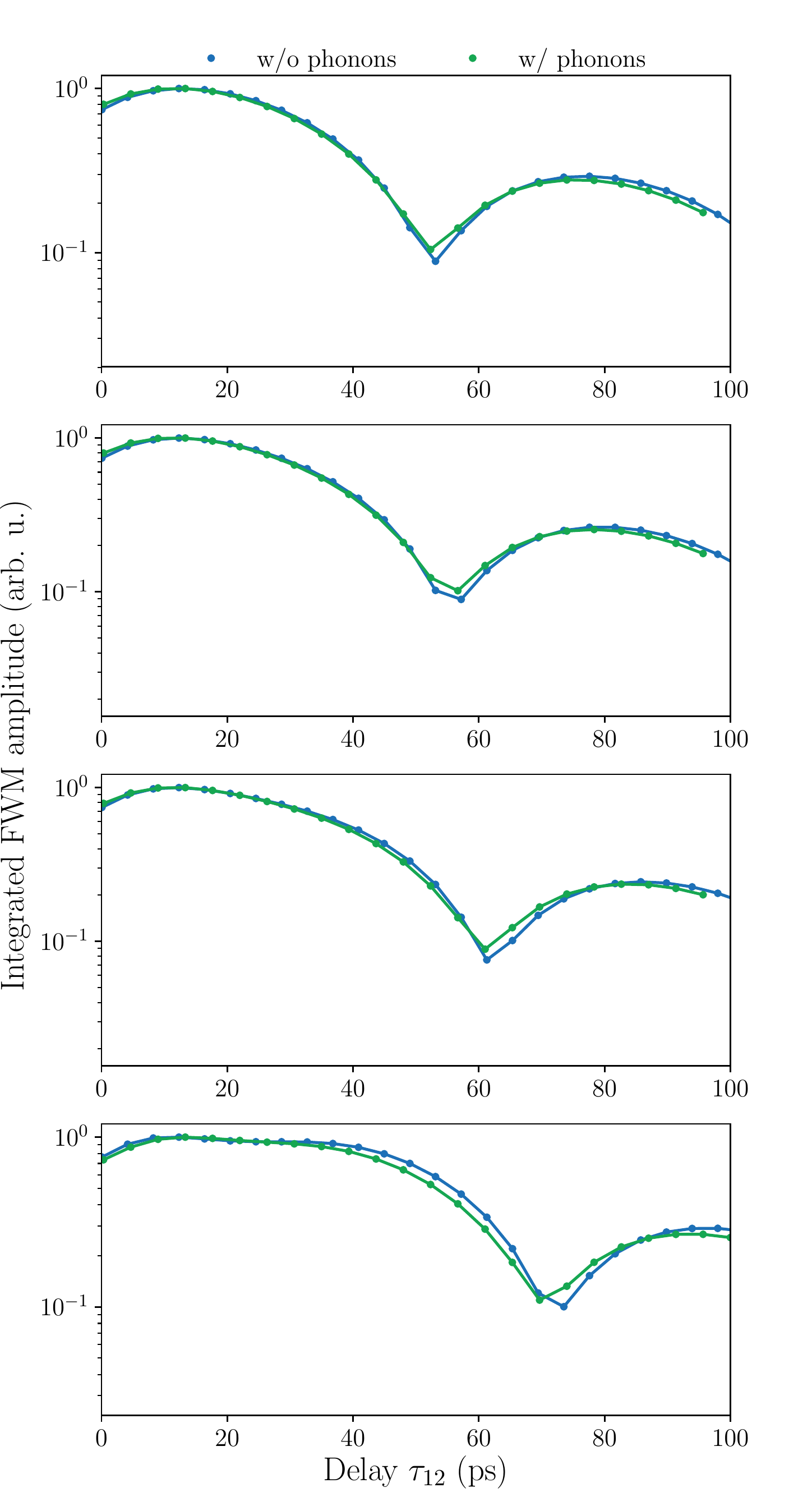}
	\caption{Delay dynamics of the integrated FWM amplitude for positive delay $\tau_{12}>0$, as already displayed in Fig.~\ref{fig:8}. Additionally to the simulation without the polaron-phonon interaction (blue dots and line), the simulated delay dynamics is shown for the first 100~ps, when taking this interaction into account (green dots and line).}
	\label{fig:12}
\end{figure}

    However, the dissipation of the cavity, described by $\mathcal{D}_a$ in Eq.~\eqref{eq:dissipator}, can secure the validity of the secular approximation even for a large initial number of photons present in the cavity. It reduces the number of relevant JC ladder rungs on a time scale that depends on the decay rate $\gamma_a$. The parts of the reduced density matrix with a large photon content decay faster for this kind of dissipator. This property originates from the fact that for an $n$-photon state it is $\bra{n-1}a\ket{n}=\sqrt{n}$, i.e. the transition amplitude to a state with less photons scales with the number of photons present in the state.

    In Fig. \ref{fig:11}(c) we check the validity of the secular approximation by comparing full TCL2 calculations (solid lines) with Lindblad calculations (dashed lines) in the case of a single pulse excitation. The absolute cavity field $|\!\braket{a}\!|$ after a single pulse is plotted as a function of time after the pulse for typical parameters, as given in the caption, and two different pulse areas $\Theta=\pi$ (red) and $\Theta=2\pi$ (blue). As expected, for the smaller pulse area the secular approximation only leads to a small deviation, whereas for the larger pulse area the deviations are more pronounced at small times around $t=20$~ps. For larger times, both the TCL2 and the Lindblad calculation agree quite well again. We conclude, that the larger the photon content in the cavity, the less accurate the secular approximation becomes. However, even for a $2\pi$ pulse we do not find a significant qualitative difference. Therefore, to render the calculations numerically feasible, we will use this approximation, keeping in mind that it might lead to small deviations of the computed FWM signals from TCL2 simulations without this approximation on short time scales.
    \section{FWM delay dynamics with phonons}
	To estimate the influence of the interaction between the polaron and the phonons on the FWM delay dynamics, Fig.~\ref{fig:12} compares simulations including this interaction (green dots and line) with simulations neglecting this interaction (blue dots and line). The situation is the same as in Fig.~\ref{fig:8}. When speaking of the polaron-phonon interaction, we mean the phonon dissipator and the Lamb shift Hamiltonian of Eqs.~\eqref{eq:Lindblad_full} and~\eqref{eq:Lindblad_full_appendix}. As can be seen in Fig.~\ref{fig:12}, these two parts of our equation of motion do not strongly influence the delay dynamics.

\end{document}